\documentclass[
11pt,
superscriptaddress,
showpacs, showkeys,
amsmath,amssymb,
aps,
pre,
floatfix,
]{revtex4-1}

\usepackage[T1]{fontenc}
\usepackage{ae,aecompl}
\usepackage{lmodern}
\usepackage{hyperref}
\usepackage{amsmath}
\usepackage{amsfonts}
\usepackage{amssymb}
\usepackage{bm}
\usepackage{xfrac}
\usepackage{graphicx}
\usepackage{siunitx}
\usepackage{etoolbox}
\usepackage{color}
\usepackage{multirow}
\usepackage{enumitem}
\usepackage{algorithm}
\usepackage{array}
\usepackage{colortbl}
\usepackage[overload]{empheq}
\usepackage[subnum]{cases}
\usepackage{xcolor}
\usepackage[normalem]{ulem}

\makeatletter
\newcommand*{\centerfloat}{%
  \parindent \z@
  \leftskip \z@ \@plus 1fil \@minus \textwidth
  \rightskip\leftskip
  \parfillskip \z@skip}
\makeatother

\setlist[itemize]{noitemsep, topsep=0pt}
\setlist[enumerate]{noitemsep, topsep=0pt}

\global\long\def\V#1{\boldsymbol{#1}}
\global\long\def\M#1{\boldsymbol{#1}}

\global\long\def\D#1{\Delta#1}
\global\long\def\d#1{\delta#1}

\global\long\def\grad{\M{\nabla}}

\global\long\def\myhalf{\sfrac{1}{2}}

\newcommand{\lapl}{{\nabla^2}}                            % discrete Laplacian
\newcommand{\divg}{{\V\nabla\cdot}}                       % divergence
\newcommand{\pddt}{{\frac{\partial}{\partial t}}}         % (partial d)/(partial dt)
\newcommand{\paren}[1]{{(#1)}}                            % (#1)
\newcommand{\chisub}{\chi^\mathrm{sub}}
\newcommand{\chisubb}{\M{\chi}^\mathrm{sub}}

\begin{document}

\title{Fluctuating Hydrodynamics of Reactive Liquid Mixtures}

\author{Changho Kim}
\email{ckim103@ucmerced.edu}
\affiliation{Computational Research Division, Lawrence Berkeley National Laboratory \\
 1 Cyclotron Road, Berkeley, CA 94720, USA }
\affiliation{Applied Mathematics, University of California, Merced \\ 5200 North Lake Road, Merced, CA 95343, USA}

\author{Andy Nonaka}
\affiliation{Computational Research Division, Lawrence Berkeley National Laboratory \\
 1 Cyclotron Road, Berkeley, CA 94720, USA }

\author{John B.\ Bell}
\affiliation{Computational Research Division, Lawrence Berkeley National Laboratory \\
 1 Cyclotron Road, Berkeley, CA 94720, USA }

\author{Alejandro L.\ Garcia}
\affiliation{Department of Physics and Astronomy, San Jose State University \\
 1 Washington Square, San Jose, CA 95192, USA }
 
\author{Aleksandar Donev}
\affiliation{Courant Institute of Mathematical Sciences, New York University \\
 251 Mercer Street, New York, NY 10012, USA }

\begin{abstract}
Fluctuating hydrodynamics (FHD) provides a framework for modeling microscopic fluctuations in a manner consistent with statistical mechanics and nonequilibrium thermodynamics.
This paper presents an FHD formulation for isothermal reactive incompressible liquid mixtures with stochastic chemistry.
Fluctuating multispecies mass diffusion is formulated using a Maxwell--Stefan description without assuming a dilute solution, and momentum dynamics is described by a stochastic Navier--Stokes equation for the fluid velocity.
We consider a thermodynamically consistent generalization for the law of mass action for non-dilute mixtures and use it in the chemical master equation (CME) to model reactions as a Poisson process.
The FHD approach provides remarkable computational efficiency over traditional reaction-diffusion master equation methods when the number of reactive molecules is large, while also retaining accuracy even when there are as few as ten reactive molecules per hydrodynamic cell.
We present a numerical algorithm to solve the coupled FHD and CME equations and validate it on both equilibrium and nonequilibrium problems.
We simulate a diffusively-driven gravitational instability in the presence of an acid-base neutralization reaction, starting from a perfectly flat interface.
We demonstrate that the coupling between velocity and concentration fluctuations dominate the initial growth of the instability.
\end{abstract}

\date{\today}

\maketitle

% spacing around equations
\setlength{\abovedisplayskip}{4pt}
\setlength{\belowdisplayskip}{8pt}
\setlength{\abovedisplayshortskip}{2pt}
\setlength{\belowdisplayshortskip}{4pt}

\section{Introduction}

Thermal fluctuations in fluids arise from random molecular motions, driving both microscopic and macroscopic behavior that deterministic models fail to predict.
In diffusive mixing experiments, velocity fluctuations lead to giant fluctuations in concentration in the presence of concentration gradients~\cite{VailatiGiglio1997}.
Buoyancy-driven instabilities can be triggered or affected by thermal fluctuations~\cite{LemaigreBudroniRiolfoGrosfilsWit2013, DonevNonakaBhattacharjeeGarciaBell2015}.
In reaction-diffusion systems, thermal fluctuations can accelerate the formation of Turing patterns on a macroscopic time scale~\cite{KimNonakaBellGarciaDonev2017}, and induce long-time memory in the chemical kinetics of a diffusion-limited system~\cite{DonevYangKim2018}.

In this paper, we develop a formulation and numerical methodology for the stochastic simulation of reactive microfluids.
Here we incorporate a stochastic description of chemical reactions based on the chemical master equation (CME)~\cite{Kampen1983} into an isothermal fluctuating hydrodynamics (FHD)~\cite{LandauLifshitz1959, ZarateSengers2006} description of diffusive and advective mass transport.
Hence, our proposed algorithm combines discrete processes (CME for reactive processes) and continuous processes (FHD for transport processes); a similar idea has been used for the simulation of the Boltzmann equation~\cite{Bird1994, MansourBaras1992, BarasMansour1997}.
The use of the CME enables us to correctly capture large fluctuations of composition, going beyond the Gaussian approximation inherent in the chemical Langevin equation (CLE) used in our prior work~\cite{BhattacharjeeBalakrishnanGarciaBellDonev2015}.
While our previous work on reaction-diffusion systems~\cite{KimNonakaBellGarciaDonev2017} also employed the CME, it was restricted to dilute solutions.
Here we generalize the CME to non-dilute ideal mixtures with a complete Maxwell--Stefan formulation of diffusive transport in multispecies mixtures.
This includes cross-diffusion coupling among distinct species and can account for deviations from ideality, unlike the standard reaction-diffusion master equation (RDME) approach~\cite{Gardiner1985, BarasMansour1996, NicolisPrigogine1977}.
Finally, by including the fluctuating Navier--Stokes equations in the model we account for advection by thermal velocity fluctuations, which is necessary to capture giant nonequilibrium composition fluctuations~\cite{MansourBaras1992, BarasMansour1997, DziekanSignonNowakowskiLemarchand2013}.

Our approach is related to, but also distinct from, prior work on fluctuating hydrodynamics for reactive liquid mixtures.
An alternative Langevin-based approach proposed in \cite{PagonabarragaPerezMadridRubi1997}, and extended to full hydrodynamics in \cite{BedeauxPagonabarragaZarateSengersKjelstrup2010}, represents reactions as a diffusion process along an internal reaction coordinate, driven by Gaussian noise.
This description is fully consistent with nonequilibrium thermodynamics and fluctuating hydrodynamics, but is not easily extensible to multispecies mixtures, and, importantly, is expensive to use in numerical simulations because it requires introducing an additional reaction coordinate, thus effectively increasing the dimensionality of the problem.
Instead, in our approach we only consider the reactant and product states and consider reactions as a jump process between these two states, driven by Poisson noise.
The deterministic (macroscopic) as well as a linearized version of the FHD equations we consider here are the same as those obtained from a quasi-stationary approximation of the model developed in \cite{BedeauxPagonabarragaZarateSengersKjelstrup2010} (see Eq.~(26) in \cite{BedeauxPagonabarragaZarateSengersKjelstrup2010} and the book by Keizer~\cite{Keizer1987}) as we have discussed in more detail in prior work~\cite{BhattacharjeeBalakrishnanGarciaBellDonev2015}.
The key difference is that here we describe chemical fluctuations using a \emph{nonlinear} FHD description based on a master equation, rather than a linearized Langevin description.
This is common in stochastic reaction-diffusion models used in biochemical modeling~\cite{GillespieHellanderPetzold2013, ErbanChapmanMaini2007}, as we have discussed in more detail in prior work~\cite{KimNonakaBellGarciaDonev2017}.
However, traditional RDME descriptions have been restricted to dilute solutions and do not account for velocity (momentum) fluctuations.
More broadly, biochemical reaction-diffusion models have largely been developed without input from the field of (non)equilibrium thermodynamics, and especially fluctuating hydrodynamics.
Here we bridge this gap by combining features of the RDME with FHD, thus delivering on the promise made in \cite{KimNonakaBellGarciaDonev2017} to ``explore combining Langevin and CME approaches together, thus further bridging the apparent gap between the two.''
Giant nonequilibrium fluctuations, which arise due to the coupling with velocity fluctuations, have been studied theoretically using linearized FHD for a dimerization reaction in~\cite{ZarateSengersBedeauxKjelstrup2007, BedeauxZaratePagonabarragaSengersKjelstrup2011}.
Here we study giant fluctuations in a liquid mixture undergoing a dimerization reaction numerically, and show that a quantitatively-accurate theoretical description is difficult due to the nonlinearity of the macroscopic steady state.

In this work we simplify our previous variable-density low Mach FHD formulation by restricting it to miscible liquid mixtures~\cite{DonevNonakaBhattacharjeeGarciaBell2015} in which the density is essentially independent of composition at fixed pressure and temperature.
The resulting Boussinesq (incompressible) approximation of the momentum equation enables us to construct an efficient numerical method that accounts for inertial effects important in buoyancy-driven fluid flows, yet remains robust for small Reynolds numbers and large Schmidt numbers.
The spatio-temporal discretization of the FHD equations is based on our previous work~\cite{DonevNonakaBhattacharjeeGarciaBell2015} but with some important improvements necessary for simulating complex reactive mixtures at small length scales. Notably, we extend our previous work on reaction-diffusion systems~\cite{KimNonakaBellGarciaDonev2017} to general multispecies mixtures so that large deviations of composition are handled accurately and robustly, and negative densities are avoided.

We follow a general framework for the systematic construction of FHD numerical methods based on the stochastic version of the method of lines approach~\cite{DonevVandenEijndenGarciaBell2010}.
Using this framework, we have previously developed stochastic simulation methods for gas mixtures~\cite{BalakrishnanGarciaDonevBell2014} and quasi-incompressible miscible liquid mixtures~\cite{DonevNonakaBhattacharjeeGarciaBell2015, NonakaSunBellDonev2015, DonevNonakaSunFaiGarciaBell2014}.
For liquid mixtures, we have developed a computationally efficient low Mach number model that eliminates fast pressure waves while preserving the spatio-temporal spectrum of the slower diffusive fluctuations~\cite{DonevNonakaSunFaiGarciaBell2014}.
To avoid severe restriction on time step size when the Schmidt number is large, we have developed an implicit temporal discretization of viscous dissipation~\cite{NonakaSunBellDonev2015} that relies on a variable-coefficient multigrid precondition to solve the coupled velocity-pressure Stokes system~\cite{CaiNonakaBellGriffithDonev2014}.

In this paper we make three novel contributions to the numerical methodology developed in our prior work.
First, by incorporating a second-order midpoint tau-leaping scheme~\cite{HuLiMin2011} into our prior algorithms for multispecies miscible liquid mixtures~\cite{DonevNonakaBhattacharjeeGarciaBell2015}, we construct a numerical method that efficiently samples reactions at a cost no larger than that of integrating the chemical Langevin equation. 
Because our novel midpoint temporal integrator solves the CME by using tau leaping~\cite{Gillespie2001}, it is robust for large composition fluctuations, while also being efficient for weak fluctuations.
Second, the midpoint scheme is constructed to be robust for large Schmidt numbers, i.e., much faster momentum diffusion compared to mass diffusion, as is typical in liquid systems. 
In particular, the numerical method reproduces the correct spectrum of giant nonequilibrium fluctuations even for time step sizes much larger than the stability limit dictated by fast momentum diffusion, while also preserving the slow inertial momentum dynamics at large scales.
Third, we take careful attention to handling vanishing species robustly both in the formulation of the multispecies diffusion model and in the numerical algorithm.

The rest of the paper is organized as follows.
In Section~\ref{sec_formul}, we present the formulation of the FHD equations coupled with the CME formulation of reactions.
In Section~\ref{sec_numerical}, we present a numerical scheme that can solve these equations accurately and robustly even in the presence of large composition fluctuations and vanishing species.
In Section~\ref{sec_examples}, we present numerical results for four examples and discuss various aspects of our numerical method and the effects of thermal fluctuations. 
First, we verify that for dilute solutions our algorithm preserves the robustness and accuracy properties of our previous method for reaction-diffusion systems~\cite{KimNonakaBellGarciaDonev2017} by modeling the hydrolysis of sucrose at micrometer scales.
Second, to assess the fidelity of our approach in a non-dilute setting, we consider a binary mixture undergoing a dimerization reaction $2\mathrm{A}\rightleftharpoons \mathrm{A}_2$ at thermodynamic equilibrium with a small number of molecules per cell.
Third, we also study such a mixture out of equilibrium in the presence of giant nonequilibrium fluctuations with a large number of molecules per cell.
Fourth, we use our numerical algorithm to simulate a diffusively-driven gravitational instability in the presence of an acid-base neutralization reaction recently studied experimentally~\cite{LemaigreBudroniRiolfoGrosfilsWit2013}, and show that the coupling between velocity and concentration fluctuations triggers and drives the instability at early times.
In Section~\ref{sec_concl}, we conclude the paper with a brief summary and a discussion of future directions.

\section{\label{sec_formul}Reactive Fluctuating Hydrodynamics}

Our formulation relies on several approximations appropriate for many isothermal miscible liquid mixtures.
First, we neglect the effects of thermodiffusion and barodiffusion on mass transport and assume constant temperature $T$ and thermodynamic pressure $P$.
Second, we assume that density variations due to composition are small enough that they have no effect on the flow field except through a buoyancy force.
Hence, we formulate our FHD system as an isothermal Boussinesq simplification of the low Mach number multispecies model used in \cite{DonevNonakaBhattacharjeeGarciaBell2015}.
While a numerical method can be potentially constructed without these approximations, the Boussinesq formulation greatly reduces the complexity of the numerical scheme without losing essential physics.

Given these approximations, we recast the continuity equation for mass density as a divergence-free constraint on velocity and assume a constant density $\rho_0$,
\begin{align}
\label{ddtvel}
\rho_0 \frac{\partial\V{v}}{\partial t} + \grad\pi &= - \rho_0 \divg(\V{v}\V{v}^\mathrm{T}) + \divg(\eta\bar{\grad}\V{v}+\M{\Sigma}) + \V{f}, \\
\label{divvel}
\divg\V{v} &= 0, \\
\label{ddtrhos}
\rho_0 \frac{\partial w_s}{\partial t} &= - \rho_0 \divg(w_s \V{v}) - \divg\V{F}_s + m_s\Omega_s.
\end{align}
Here, $\V{v}$ is the fluid velocity, $\pi$ is the mechanical pressure (a Lagrange multiplier that ensures the velocity remains divergence free~\cite{ChorinMarsden1990}), $\eta(\V{w})$ is the viscosity, $\bar{\grad}=\grad+\grad^\mathrm{T}$ is a symmetric gradient, and $\M{\Sigma}$ is the stochastic momentum flux.
By denoting the number of species with $N_\mathrm{spec}$, the vector of mass fractions (concentrations) is given by $\V{w}=(w_1,\dots,w_{N_\mathrm{spec}}$), where $w_s$ is the mass fraction of species $s$ and $\sum_s w_s = 1$.
We compute the mass density of each species using $\rho_s = \rho_0 w_s$ and thus the total mass density $\sum_s \rho_s = \rho_0$ is strictly constant.
The buoyancy force $\V{f}(\V{w})$ is a problem-specific function of $\V{w}$.
The total diffusive mass flux  $\V{F}_s$ of species $s$ is decomposed into a dissipative flux $\overline{\V{F}}_s$ and fluctuating flux $\widetilde{\V{F}}_s$,
\begin{equation}
\V{F}_s = \overline{\V{F}}_s + \widetilde{\V{F}}_s,
\end{equation}
and $m_s\Omega_s$ represents a source term representing stochastic chemistry, where $m_s$ is the molecular mass and $\Omega_s$ is the number density production rate for species $s$.
Note that by summing up \eqref{ddtrhos} over all species we recover \eqref{divvel} since $\sum_s\V{F}_s=\V{0}$ and $\sum_s m_s\Omega_s=0$.
Based on the fluctuation-dissipation relation, the stochastic momentum flux $\M{\Sigma}$ is modeled as
\begin{equation}
\M{\Sigma} = \sqrt{\eta k_\mathrm{B} T} \left[ \M{\mathcal{Z}}^\text{mom} + (\M{\mathcal{Z}}^\text{mom})^\mathrm{T} \right],
\end{equation}
where $k_\mathrm{B}$ is Boltzmann's constant, and $\M{\mathcal{Z}}^\text{mom}(\V{r},t)$ is a standard Gaussian white noise (GWN) tensor field with uncorrelated components having $\delta$-function correlations in space and time.

We formulate multispecies diffusion in Section~\ref{subsec_MSdiff}, and chemistry in Section~\ref{subsec_chem}.
It is important to note that both the diffusion and chemistry formulations are obtained from a general form of the specific chemical potential for each species,
\begin{equation}
\label{mus}
\mu_s(\V{x},T,P) = \mu_s^0(T,P) + \frac{k_\mathrm{B}T}{m_s}\log (x_s\gamma_s),
\end{equation}
where $\mu_s^0(T,P)$ is a reference chemical potential and $\gamma_s(\V{x},T,P)$ is the activity coefficient (for an ideal mixture, $\gamma_s=1$).
Here $\V{x}$ denotes mole fractions, which can be expressed in terms of $\V{w}$ as
\begin{equation}
\label{w2x}
\V{x} = \bar{m}\left(\frac{w_1}{m_1},\dots,\frac{w_{N_\mathrm{spec}}}{m_{N_\mathrm{spec}}}\right),
\end{equation}
where $\bar{m}$ is the mixture-averaged molecular mass,
\begin{equation}
\label{barm}
\bar{m} = \left(\sum_s \frac{w_s}{m_s}\right)^{-1}.
\end{equation}
In Section~\ref{subsec_formul_dimer}, we confirm the thermodynamic consistency of our formulation by showing that thermodynamic equilibrium is determined by the chemical potentials, and that transport processes and reactions do \emph{not} change equilibrium statistics.
In Section~\ref{subsec_formul_dilite}, we discuss the simplification of our model for dilute solutions.

\subsection{\label{subsec_MSdiff}Multispecies Diffusion}

Here we summarize the FHD description of multispecies diffusion formulated in \cite{DonevNonakaBhattacharjeeGarciaBell2015}.
Neglecting thermodiffusion and barodiffusion, the Maxwell--Stefan formulation of the diffusion driving force gives
\begin{equation}
\label{drivforce}
\M{\Gamma}\grad\V{x} = -\rho_0^{-1}\M{\Lambda}\M{W}^{-1}\overline{\M{F}},
\end{equation}
where $\M{\Gamma}$ is the matrix of thermodynamic factors that becomes the identity matrix for ideal mixtures, and $\M{W}$ is a diagonal matrix with entries $\V{w}$.
The symmetric matrix $\M{\Lambda}$ is defined via 
\begin{equation}
\label{Lambda}
\Lambda_{ss'} = - \frac{x_s x_{s'}}{\textsc{\DJ}_{ss'}} \mbox{ if $s\ne s'$ and }\Lambda_{ss} = -\sum_{s'\ne s} \Lambda_{ss'},
\end{equation}
where $\textsc{\DJ}_{ss'}$ is the Maxwell--Stefan binary diffusion coefficient between species $s$ and $s'$.
Denoting a pseudo-inverse of $\M{\Lambda}$ with $\M{\chi}$, we can rewrite \eqref{drivforce} as
\begin{equation}
\label{detmassflux}
\overline{\M{F}} = -\rho_0\M{W}\M{\chi}\M{\Gamma}\grad{\V{x}}.
\end{equation}

The stochastic mass fluxes $\widetilde{\M{F}}$ are given by the fluctuation-dissipation relation,
\begin{equation}
\label{stochmassflux}
\widetilde{\M{F}} = \sqrt{2\bar{m}\rho_0} \; \M{W}\M{\chi}^\frac{1}{2}\V{\mathcal{Z}}^\text{mass},
\end{equation}
where $\M{\chi}^\frac{1}{2}$ is a ``square root'' of $\M{\chi}$ satisfying $\M{\chi}^\frac12 (\M{\chi}^{\frac12})^\mathrm{T} = \M{\chi}$, and $\M{\mathcal{Z}}^\text{mass}(\V{r},t)$ is a standard GWN field with uncorrelated components.
Modifications of this formulation in the presence of trace or vanishing species are discussed in Section~\ref{subsec_spatial}.

\subsection{\label{subsec_chem}Chemical Reactions}

We consider a liquid mixture undergoing $N_\mathrm{react}$ elementary reversible reactions of the form 
\begin{equation}
\sum_{s=1}^{N_\mathrm{spec}} \nu_{sr}^+\mathfrak{M}_s \rightleftharpoons \sum_{s=1}^{N_\mathrm{spec}} \nu_{sr}^-\mathfrak{M}_s \quad (r=1,\dots,N_\mathrm{react}),
\end{equation}
where $\nu^\pm_{sr}$ are molecule numbers, and $\mathfrak{M}_s$ are chemical symbols.
We define the stoichiometric coefficient of species $s$ in the forward reaction $r$ as $\D{\nu}^+_{sr}=\nu_{sr}^- - \nu_{sr}^+$ and the coefficient in the reverse reaction as $\D{\nu}^-_{sr} = \nu^+_{sr}-\nu^-_{sr}$.
We assume that mass conservation holds in each reaction $r$; i.e., $\sum_s \D{\nu}^\pm_{sr} m_s = 0$ for all $r$.
It is important to note that all reactions must be reversible for thermodynamic consistency.

To sample $\Omega_s$, we need propensity density functions $a^\pm_r$ for the forward/reverse ($+/-$) rates of reaction $r$.
Specifically, the mean number of reaction occurrences in a locally well-mixed reactive cell of volume $\D{V}$ during an infinitesimal time interval $dt$ is given as $a^\pm_r\D{V}dt$.
Accordingly, the mean number density production rate of species $s$ is given as
\begin{equation}
\overline{\Omega}_s = \sum_r\sum_{\alpha=\pm} \D{\nu}^\alpha_{sr} a^\alpha_r.
\end{equation}
In Section~\ref{subsubsec_genLMA} we give a generalized law of mass action (LMA) based on thermodynamically consistent $a^\pm_r$, and in Section~\ref{subsubsec_CMEstochchem} we present a CME-based stochastic formulation of chemical reactions.

\subsubsection{\label{subsubsec_genLMA}Generalized Law of Mass Action}

Here we adopt the canonical form for the rate of chemical reactions~\cite{Marcelin1910, Keizer1987}.
Propensity density functions are expressed as~\cite{BhattacharjeeBalakrishnanGarciaBellDonev2015} 
\begin{equation}
a_r^\pm = \lambda_r\prod_s e^{\nu_{sr}^\pm\hat{\mu}_s},
\end{equation}
where $\lambda_r (T,P)\ge 0$ is a reaction rate parameter assumed to be independent of the composition, and $\hat{\mu}_s = m_s\mu_s / k_\mathrm{B}T$ is the dimensionless chemical potential per particle.
For the general form of chemical potential~\eqref{mus}, we have
\begin{equation}
\label{mfLMA}
a_r^\pm = \kappa_r^\pm \prod_s (x_s \gamma_s)^{\nu^\pm_{sr}},
\end{equation}
where $\kappa^\pm_r(T,P) = \lambda_r\prod_s\exp(\nu_{sr}^\pm \hat{\mu}_s^0)$ denotes the forward/reverse reaction rate constant.
From the condition $a_r^+=a_r^-$ at chemical equilibrium, we can express the equilibrium constant as a purely thermodynamic quantity, 
\begin{equation}
\label{eqconst}
K_r(T,P) = \frac{\kappa^+}{\kappa^-} = \exp\left(-\sum_s \D{\nu}^+_{sr}\hat{\mu}_s^0\right), 
\end{equation}
as required by statistical mechanics.

It is important to note that propensity density functions and equilibrium constants are expressed in terms of \emph{mole fractions} $x_s$ (for ideal mixtures) or activities $x_s\gamma_s$.
This generalized LMA has a different form compared to the \emph{number density} based LMA used for ideal gas mixtures in our prior work~\cite{BhattacharjeeBalakrishnanGarciaBellDonev2015}.
However, this does not imply any incompatibility between the two forms of LMA.
For isothermal gas mixtures, pressure changes significantly upon reaction due to changes in mole numbers and thus $\kappa^\pm_r(T,P)$ cannot be assumed to be constant.
On the other hand, in liquid mixtures, where pressure changes are not significant, $\kappa^\pm_r(T,P)$ can be assumed to be constant.

\subsubsection{\label{subsubsec_CMEstochchem}CME-based Stochastic Chemistry}

We believe that an accurate mesoscopic chemistry description should be based on a master equation approach, which leads to the CME~\cite{GillespieHellanderPetzold2013} for well-mixed~\footnote{
In a reaction-diffusion setting, this means that diffusion dominates on the length scale $\D{x}$ of a reactive cell.
Equivalently, for typical diffusion coefficient $D$ and linearized reaction rate $r$, the penetration depth $\xi=\sqrt{D/r}$ is significantly larger than $\D{x}$, which is itself much larger than molecular scales.
In this reaction-limited case, the validity of the mesoscopic description of reactions is guaranteed.
On the other hand, for a diffusion-limited system, where $\xi$ becomes comparable to molecular scales, the validity of the CME remains to be investigated~\cite{DonevYangKim2018}.
} % for footnote
systems.
As will be demonstrated in Section~\ref{subsubsec_formul_dimer_eq_dist}, both the CME description and the generalized LMA are crucial for achieving thermodynamic consistency.
Note, however, that our CME-based description itself does not require reversible reactions.
For modeling purposes, one can exclude some forward or reverse reactions by assuming they have zero rates. 
However, we remind the reader that this is inconsistent with equilibrium thermodynamics.

For reactions in a closed well-mixed
cell of volume $\D{V}$, the CME describes the time evolution of the system in terms of the temporal change in the \emph{probability} of the system to occupy each state (specified by the number of molecules of each species).
We use an equivalent, but more direct, \emph{trajectory-wise} representation~\cite{KimNonakaBellGarciaDonev2017}, which is related to the computationally efficient tau leaping method~\cite{Gillespie2001}.
The change in the number of molecules $N_s$ of species $s$ in a given cell during an infinitesimal time interval $dt$ is expressed in terms of the number of occurrences $\mathcal{P}(a_r^\pm\D{V}dt)$ of each reaction $r$,
\begin{equation}
\label{Rs}
{d N_s} = \Omega_s \D{V} dt = \sum_r\sum_{\alpha=\pm} \D{\nu}^\alpha_{sr} {\mathcal{P} (a_r^\alpha \D{V}dt)},
\end{equation}
where $\mathcal{P}(m)$ denotes a Poisson random variables with mean $m$.
Note that the instantaneous rate of change is written as an Ito stochastic term. The tau leaping method discretizes \eqref{Rs} with a finite time step size $\D{t}$.
To faithfully model the discrete nature of reactions, we sample \emph{integer-valued} reaction counts using Poisson random numbers as in the traditional tau leaping algorithm. However, it is important to note that we use \emph{continuous-ranged} number densities for advection-diffusion, and therefore cells are not guaranteed to have an integer number of molecules.

We note that a Gaussian approximation of the Poisson random number $\mathcal{P}(a_r^\pm\D{V}dt)$ in \eqref{Rs} leads to the chemical Langevin equation (CLE).
In this Langevin (Gaussian noise) approximation~\cite{BhattacharjeeBalakrishnanGarciaBellDonev2015, KimNonakaBellGarciaDonev2017},
\begin{equation}
\label{OmegaLangevin}
\Omega_s^\mathrm{CLE} = \sum_r\sum_{\alpha=\pm} \D{\nu}^\alpha_{sr}\left(a_r^\alpha + \sqrt{a_r^\alpha}\mathcal{Z}^\text{react}_r\right),
\end{equation}
where $\mathcal{Z}^\text{react}_r(\V{r},t)$ denotes a standard GWN field.
The Langevin description is justified in the limit of small Gaussian fluctuations with respect to average concentrations~\cite{GillespieHellanderPetzold2013}.
However, the Langevin description predicts an unphysical equilibrium state with negative densities, and does not correctly model large deviations of chemical fluctuations~\cite{BhattacharjeeBalakrishnanGarciaBellDonev2015}.
By contrast, the tau leaping method correctly reproduces the large deviation functional of the CME, while still being to remaining as efficient as the CLE (see discussion around Eq.~(8.1) in \cite{KellyVandenEijnden2017}).

\subsection{\label{subsec_formul_dimer}Thermodynamic Consistency}

We now demonstrate the thermodynamic consistency of our formulation for ideal mixtures at thermodynamic equilibrium.
For the simplicity of exposition, we consider a binary liquid mixture of $\mathrm{A}$ atoms and $\mathrm{A}_2$ molecules undergoing a dimerization reaction
\begin{equation}
\label{dimer_rxn}
2 \mathrm{A} \rightleftharpoons \mathrm{A}_2,
\end{equation}
noting that this analysis also applies to multispecies ideal mixtures.   
In Section~\ref{subsubsec_formul_dimer_eq_dist}, we consider the single-cell (homogeneous) case.
We obtain the thermodynamic equilibrium distribution of monomers and dimers to show that our chemistry model satisfies detailed balance with respect to the correct Einstein equilibrium distribution.
In Section~\ref{subsubsec_formul_dimer_struct_fac}, we consider the spatially extended case. 
We show that the governing Boussinesq equations give flat structure factors at thermodynamic equilibrium in the Gaussian approximation, in agreement with statistical mechanics.

\subsubsection{\label{subsubsec_formul_dimer_eq_dist}Single-Cell System}

We denote the number of monomers and dimers as $N_1$ and $N_2$, respectively.
By the constant density approximation, $N_1$ and $N_2$ satisfy $N_1 + 2 N_2 = \rho_0\D{V}/m \equiv N_0$, where $m$ is the mass of a monomer and $N_0$ is the total number of $\mathrm{A}$ atoms in a cell of volume $\D{V}$.
Hence, we denote the equilibrium distribution of the composition with $P(N_2)$.

Statistical mechanics predicts that the equilibrium distribution is given by the Einstein distribution, $P \sim e^{S/k_\mathrm{B}}$, where $S$ denotes the entropy of the system at a given state.
Note that even though we consider isothermal systems, we can still use the Einstein distribution since the only contribution to the free energy that depends on composition is the entropy of mixing.
For a binary ideal mixture, the entropy of mixing is given as
\begin{equation}
\label{dSmix}
S_\mathrm{mix}(N_1,N_2) = k_\mathrm{B} \log \frac{(N_1+N_2)!}{N_1! N_2!},
\end{equation}
and the entropy of the system is
\begin{equation}
\label{SN1N2}
S(N_1,N_2) = S_\mathrm{mix}(N_1,N_2) - k_\mathrm{B} ( N_1\hat{\mu}_1^0 + N_2\hat{\mu}_2^0).
\end{equation}
Hence, we obtain the equilibrium distribution
\begin{equation}
\label{eqdist}
P(N_2)\sim e^{S(N_0-2N_2,N_2)/k_\mathrm{B}}
\end{equation}
with $\sum_{N_2=0}^{N_0/2}P(N_2)=1$.
Note that it is straightforward to obtain the ratio of occupation probabilities of adjacent states,
\begin{equation}
\label{probrat}
\frac{P(N_2+1)}{P(N_2)} = \frac{(N_0-2N_2)(N_0-2N_2-1)}{(N_0-N_2)(N_2+1)} \exp(2\hat{\mu}_1^0-\hat{\mu}_2^0).
\end{equation}

We now analyze when detailed balance is achieved for the dimerization reaction~\eqref{dimer_rxn} with respect to the equilibrium distribution $P(N_2)$.
The detailed balance condition is given as
\begin{equation}
\label{detbal}
P(N_2) a^+(N_2) = P(N_2+1) a^- (N_2+1),
\end{equation}
where $a^\pm(N_2)$ denote the forward/reverse rates at the state with $N_2$ dimers, which are to be determined.
By using \eqref{eqconst} and \eqref{probrat}, one can show that the detailed balance condition~\eqref{detbal} exactly holds for
\begin{subequations}
\label{corrLMA}
\begin{align}
\label{corrLMA1}
& a^+ (N_2) \equiv \kappa^+ \left(\frac{N_1}{N_1 + N_2}\right) \left(\frac{N_1-1}{N_1 + N_2 - 1}\right),\\
\label{corrLMA2}
& a^- (N_2) \equiv \kappa^- \left(\frac{N_2}{N_1+N_2}\right),
\end{align}
\end{subequations}
with $N_1 = N_0 - 2 N_2$.
It is important to note that \eqref{corrLMA} reduces to $a^+ = \kappa^+ x_1^2$ and $a^- = \kappa^- x_2$ in the thermodynamic limit.
Hence, \eqref{corrLMA1} can be considered as an integer-based correction to the generalized LMA~\eqref{mfLMA}; this correction makes sense because the probability of choosing a second monomer is $(N_1-1)/(N_1+N_2-1)$.
Such integer corrections are well known for low density solutions and used in most RDME models of reaction-diffusion systems, but to our knowledge they have not previously been formulated for non-dilute ideal mixtures.

In the thermodynamic limit, we can apply Stirling's approximation to \eqref{dSmix} and express chemical potentials in \eqref{SN1N2} in terms of equilibrium mole fractions $x_s^\mathrm{eq}$, to give
\begin{equation}
\label{S_Stirling}
S_\mathrm{Stirling} = S^\mathrm{eq} -k_\mathrm{B} N \sum_s x_s \log(x_s/x_s^\mathrm{eq}),
\end{equation}
where $S^\mathrm{eq}$ denotes the entropy at $\V{x}^\mathrm{eq}$ and $N=\sum_s N_s$.
We can further approximate $S_\mathrm{Stirling}$ up to second order in $\d{x}_s=x_s-x_s^\mathrm{eq}$ ($s=1,\dots,N_\mathrm{spec}-1$), to get a Gaussian approximation to the Einstein distribution.
This Gaussian approximation is described by linearized FHD and we study it in more detail, including spatial dependence, next.

\subsubsection{\label{subsubsec_formul_dimer_struct_fac}Spatially Extended System}

We can extend the dimerization results obtained for the single-cell case to the spatially extended case.
For an ideal mixture the total entropy of the system is additive over the individual cells,
\begin{equation}
\label{eq:S_tot}
S_\mathrm{tot} = \sum_i S(N_0-2N_2^{(i)},N_2^{(i)}),
\end{equation}
where $N_2^{(i)}$ denotes the number of dimers in cell $i$. Therefore, the Einstein distribution for the spatially extended system is the product distribution
\begin{equation}
\label{eq:P_tot}
P_\mathrm{tot} = \prod_i P(N_2^{(i)}).
\end{equation}
This means that the number of dimers in each cell is independent of those in the other cells and has the same distribution as the single-cell case.

We note that our FHD model of multispecies diffusion is constructed so that it reproduces the correct Einstein distribution under Stirling's approximation, i.e., our model is consistent with \eqref{S_Stirling} and \eqref{eq:S_tot}.
Hence, the combined chemistry and FHD model is expected to give the correct equilibrium distribution, as long as there are sufficiently many molecules of all species in each cell to justify the continuous approximation.
In Section~\ref{subsec_dimer_dist}, we numerically confirm that our method gives an accurate approximation to \eqref{eq:P_tot} even when there are significant fluctuations of composition, with as few as $N_2\sim 10$ dimers per cell.

At the level of a Gaussian approximation, we can investigate the system analytically using the linearized FHD equations. 
We denote the mass fractions of monomers and dimers as $w$ and $1-w$, respectively.
We assume that $w$ fluctuates around $\bar{w}$.
At equilibrium, our FHD equations~\eqref{ddtvel}--\eqref{ddtrhos} are linearized for $\V{v}=\d{\V{v}}$ and $w=\bar{w}+\d{w}$ as follows:
\begin{subequations}
\begin{align}
\label{ddtv_lin}
\partial_t(\d{\V{v}}) &= -\rho_0^{-1}\grad\pi + \nu \lapl (\d{\V{v}}) +\sqrt{\nu k_\mathrm{B}T \rho_0^{-1}} \divg \left[\M{\mathcal{Z}}^\text{mom}+(\M{\mathcal{Z}}^\text{mom})^\mathrm{T}\right],\\
\label{divv_lin}
\divg (\d{\V{v}}) &= 0,\\
\label{ddtw_lin}
\partial_t(\d{w}) &= D \lapl (\d{w}) + \sqrt{2D k_\mathrm{B}T \rho_0^{-1}\mu_w^{-1}}\divg \V{\mathcal{Z}}^\text{mass} + \rho_0^{-1} m\Omega_1^\mathrm{lin},
\end{align}
\end{subequations}
where $\nu=\eta/\rho_0$, $D=\textsc{\DJ}_{12}$, and $\mu_w$ is the second order derivative of Gibbs free energy with respect to concentration $w$, given as $\mu_w = k_\mathrm{B}T/[m\bar{w}(1-\bar{w}^2)]$ for an ideal mixture.
The linearized reaction term is denoted by $\rho_0^{-1}m\Omega_1^\mathrm{lin}$.

We denote the equilibrium structure factors (spectra) by $S^\mathrm{eq}_{\V{v},\V{v}}(\V{k})=\langle \d{\hat{\V{v}}}\d{\hat{\V{v}}}^* \rangle$ and $S^\mathrm{eq}_{w,w}(\V{k})=\langle \d{\hat{w}}\d{\hat{w}}^* \rangle$, where hat denotes a Fourier transform, and asterisk denotes a conjugate transpose. 
Noting that the concentration equation is uncoupled from the momentum equation, these structure factors can be obtained separately.
For the non-reactive case, they are independent of $k$~\cite{DonevNonakaSunFaiGarciaBell2014},
\begin{equation}
\label{Seq}
S_{\V{v},\V{v}}^\mathrm{eq} = \frac{k_\mathrm{B}T}{\rho_0}\M{I},\quad
S_{w,w}^\mathrm{eq}=\frac{k_\mathrm{B}T}{\rho_0\mu_w}=\frac{m}{\rho_0}\bar{w}(1-\bar{w}^2).
\end{equation}
It is easy to show that the spatial correlations of the composition fluctuations $S_{w,w}^\mathrm{eq}$ are fully consistent with the Gaussian approximation of \eqref{eq:P_tot}.

For the reactive case, \eqref{ddtw_lin} contains the stochastic chemistry term
\begin{equation}
\label{dimer_lin_cle}
\rho_0^{-1}m\Omega_1^\mathrm{lin} = - r(\d{w}) + \sqrt{\frac{8m^2(1-\bar{w})}{\rho_0^2(1+\bar{w})}\kappa^-}\; \V{\mathcal{Z}}^\text{react},
\end{equation}
where the linearized reaction rate is
\begin{equation}
\label{dimer_lin_rate}
r = \frac{4m}{\rho_0\bar{w}(1+\bar{w})^2}\kappa^{-}.
\end{equation}
This is obtained by linearizing the Langevin expression~\eqref{OmegaLangevin}.
One can easily show that the inclusion of the reaction term does not change $S_{w,w}^\mathrm{eq}$, consistent with thermodynamic equilibrium.
This explicitly confirms that our formulation is consistent with equilibrium statistical mechanics at the level of a Gaussian approximation of the fluctuations.

\subsection{\label{subsec_formul_dilite}Dilute Limit}

One of the common assumptions in traditional reaction-diffusion modeling is that each chemical species is dilute and thus diffuses independently of other species.
In this section we explain how our formulation simplifies in the dilute limit.
We consider a solution where all solute species are dilute (i.e., $x_s\ll 1$) but the solvent is possibly a homogeneous mixture.
We use index $s$ here to denote only solute species.
In the dilute limit, $\gamma_s\rightarrow 1$ and the solute number densities are linearly proportional to their mole fractions, $n_s\approx (\rho_0/\bar{m}_\mathrm{sol}) x_s$, where $\bar{m}_\mathrm{sol}$ is the mixture-averaged molecular mass among solvent species, see \eqref{barm}.
Hence, $\hat{\mu}_s$ can be expressed in terms of $n_s$,
\begin{equation}
\hat{\mu}_s = \left(\hat{\mu}_s^0 + \log\frac{\bar{m}_\mathrm{sol}}{\rho_0}\right) + \log n_s,
\end{equation}
and consequently, the generalized (mole fraction based) LMA can be cast into the form of the traditional (number density based) LMA,
\begin{equation}
\label{ndLMA}
a_r^\pm = k_r^\pm \prod_s n_s^{\nu_{sr}^\pm},
\end{equation}
where $k_r^\pm$ denote reaction rate constants.

In the dilute limit, multispecies diffusion also becomes simpler.
In Appendix~\ref{appendix_Wchi}, we consider the dilute limit of a single solute species dissolved in a solvent mixture and show that the diffusion of the solute species is decoupled from solvent species, see \eqref{barF_N}.
It is straightforward to extend this result to multiple solute species.
The diffusion coefficient of each solute species $s$ then becomes a constant, that is, decoupled from the other species, yielding
\begin{equation}
\label{dilutelimit}
\pddt n_s = D_s \lapl n_s + \divg\left[ \sqrt{2D_s n_s}\V{\mathcal{Z}}_s\right] + \Omega_s,
\end{equation}
where $\Omega_s$ represent stochastic chemistry terms based on the LMA \eqref{ndLMA}.
Therefore, in the absence of fluid flow, our formulation is reduced to our previous reaction-diffusion model~\cite{KimNonakaBellGarciaDonev2017} in the dilute limit.
Note that jump processes and diffusion processes are combined in \eqref{dilutelimit} and the time evolution of the probability distribution of $\V{n}=\{n_s\}$ can be described by the differential Chapman--Kolmogorov equation~\cite{Gardiner1985}.

\section{\label{sec_numerical}Numerical Method}

In developing a numerical method to solve \eqref{ddtvel}--\eqref{ddtrhos}, we seek an approach that
\begin{itemize}
\item Exhibits second-order accuracy in space and time deterministically, and second-order weak accuracy in time for the linearized FHD equations~\cite{DelongSunGriffithVandenEijndenDonev2014}.
\item Reduces to our previous method for reaction-diffusion systems~\cite{KimNonakaBellGarciaDonev2017} in the dilute limit, in the absence of fluid flow.
\item Generates accurate structure factors for both equilibrium and giant fluctuations, even for large Schmidt numbers.
\item Is robust in presence of trace or vanishing species.
\end{itemize}
We explain below how our design decisions satisfy these requirements.
In Section~\ref{subsec_spatial}, we review our spatial discretization scheme and discuss robust numerical approaches for avoiding negative densities and treating vanishing species.
In Section~\ref{subsec_temporal}, we present our temporal integration scheme.
In Section~\ref{subsec_num_struct_fac}, we analyze the weak accuracy of our temporal integrator.

\subsection{\label{subsec_spatial}Spatial Discretization}

Our spatial discretization is identical to the one used in our previous work on non-reactive FHD~\cite{DonevNonakaBhattacharjeeGarciaBell2015, NonakaSunBellDonev2015, DonevNonakaSunFaiGarciaBell2014, UsabiagaBellDelgadoBuscalioniDonevFaiGriffithPeskin2012}, with a few modifications noted below.
The numerical framework is a structured-grid finite-volume approach with cell-averaged densities and pressure, and face-averaged (staggered) velocities.
We use standard second-order stencils for the gradient, divergence, and spatial averaging in order to satisfy discrete fluctuation-dissipation balance~\cite{DonevVandenEijndenGarciaBell2010}.

For the densities, we construct all mass fluxes on faces and employ the standard conservative divergence.
For the advective mass fluxes, we implement two options.
Centered advection uses two-point averaging of densities to faces, and is nondissipative and thus preserves the spectrum of fluctuations~\cite{DonevVandenEijndenGarciaBell2010}.
However, in order to prevent unphysical oscillations in mass densities in high P\'eclet number flows with sharp gradients, we can also use the Bell--Dawson--Shubin (BDS) second-order Godunov advection scheme~\cite{BellDawsonShubin1988, NonakaMayAlmgrenBell2011}.
We note that BDS advection adds artificial dissipation and does not obey a fluctuation-dissipation principle, but is necessary for simulations where centered advection would fail due to insufficient spatial resolution.
All simulations in this paper use centered advection unless otherwise noted.
The discretization of the momentum equation is the same as our previous work~\cite{DonevNonakaBhattacharjeeGarciaBell2015, NonakaSunBellDonev2015}.
We allow for periodic boundary conditions, impermeable walls, and no-flow reservoirs~\cite{DonevNonakaSunFaiGarciaBell2014, UsabiagaBellDelgadoBuscalioniDonevFaiGriffithPeskin2012} held at fixed concentrations.

The first modification relative to our previous work~\cite{DonevNonakaBhattacharjeeGarciaBell2015} is that for the stochastic mass fluxes $\widetilde{\V{F}}$, we compute the matrix $\sqrt{2\bar{m}\rho_0}\M{W}\M{\chi}^\frac{1}{2}$ directly on the face using spatially averaged densities, rather than computing this matrix at cell centers and averaging to faces.
To compute the spatial averages, we use a \emph{modified} arithmetic averaging function~\cite{KimNonakaBellGarciaDonev2017},
\begin{equation}
\label{ntilde}
\tilde{n}(n_1,n_2) = \frac{n_1+n_2}{2}H(n_1\D{V})H(n_2\D{V}),
\end{equation}
where $n_1$ and $n_2$ denote number densities at the cell centers of two neighboring cells, and $H$ is a smoothed Heaviside function defined as
\begin{equation}
H(x)=\begin{cases}
0 & \mbox{for $x\le 0$},\\
x & \mbox{for $0\le x \le 1$},\\
1 & \mbox{for $x\ge 1$}.
\end{cases}
\end{equation}
Specifically, we first convert cell-centered mass fractions to number densities, then apply $\tilde{n}$ to obtain face-centered number densities, and finally convert these back to mass fractions that are used to compute $\sqrt{\bar{m}}\M{W}\M{\chi}^{\frac12}$.
We note that $\tilde{n}$ drives the average (and thus stochastic flux) to zero if the number of molecules in either neighboring cell is sufficiently small (i.e., $n_i\D{V}\le 1$), which prevents the occurrence of negative number densities.
In most cases of interest, small numbers of molecules per cell correspond to dilute species.
For dilute species (see \eqref{dilutelimit}), the validity of using $\tilde{n}$ has been justified in \cite{KimNonakaBellGarciaDonev2017}.
In Section~\ref{subsec_dimer_dist}, we numerically confirm that our approach is robust even when the total number of molecules in a cell is $\mathcal{O}(10)$. 

Another key modification is the computation of the diffusion matrix $\M{\chi}$ for the deterministic and stochastic mass fluxes in the absence of some species, or, in the presence of vanishing species.
In the vanishing limit, where one or more concentrations become zero, the diffusion matrix $\M{\chi}$ is not well conditioned since the corresponding diagonal component $\chi_{ss}$ diverges.
This can cause numerical issues when one attempts to compute $\overline{\M{F}}$ and $\widetilde{\M{F}}$ since they depend on $\M{W}\M{\chi}$ and $\M{W}\M{\chi}^\frac{1}{2}$, respectively.
Unlike $\M{\chi}$, however, the matrices $\M{W}\M{\chi}$ and $\M{W}\M{\chi}\M{W}$ are well defined in the vanishing limit, and we can construct $\M{W}\M{\chi}$ using a special procedure.
The basic idea is that we first compute a diffusion sub-matrix $\chisubb$ of $\M{\chi}$ with the rows and columns corresponding to each vanishing species omitted.
Then we expand this sub-matrix into the full matrix $\M{W}\M{\chi}$ and approximate the remaining components using the mathematical limit of vanishing species, $w_s\rightarrow 0^+$ for all vanishing species $s$.

To formally describe the procedure for computing $\M{W}\M{\chi}$ in the vanishing limit, we introduce a mapping, $\mathfrak{m}(i)$, used to expand/contract a subsystem matrix to/from a full matrix.
For example, in a 6-species system having vanishing species $w_2$ and $w_4$, we have $\mathfrak{m}(\M{i}) = (1,0,2,0,3,4)$, $\M{i}=(1,\dots,6)$.
As graphically illustrated using the 6-species system in Fig.~\ref{fig:expansion_matrix}, there are four cases to consider when one populates $(\M{W}\M{\chi})_{ij}$:
\begin{numcases}
{\label{eq:Wchi}(\M{W}\M{\chi})_{ij}=}
\label{eq:Wchi_yellow}
w_i \chisub_{\mathfrak{m}(i)\mathfrak{m}(j)}, & $\mathfrak{m}(i) \ne 0$, $\mathfrak{m}(j) \ne 0$~({\rm yellow}), \\
\label{eq:Wchi_red}
\frac{m_i D_i}{\bar{m}}, & $\mathfrak{m}(i)=0$, $j = i$~({\rm red}), \\
\label{eq:Wchi_zero}
0, & $\mathfrak{m}(i)=0$, $j\ne i$, \\
\label{eq:Wchi_cyan}
w_i D_j \Bigg[\sum_{\substack{k\\\mathfrak{m}(k)\ne 0}} \frac{x_k}{\textsc{\DJ}_{kj}} \chisub_{\mathfrak{m}(i)\mathfrak{m}(k)} -\frac{m_j}{\bar{m}}\Bigg], & $\mathfrak{m}(i)\ne 0$, $\mathfrak{m}(j) = 0$~({\rm blue}),
\end{numcases}
where
\begin{equation}
\label{Deff}
D_j = \Bigg[ \sum_{\substack{k\\\mathfrak{m}(k)\ne 0}} \frac{x_k}{\textsc{\DJ}_{kj}} \Bigg]^{-1}.
\end{equation}
Note that color names in the parentheses in \eqref{eq:Wchi} correspond to the colors in Fig.~\ref{fig:expansion_matrix}.
A derivation of \eqref{eq:Wchi} and \eqref{Deff} is presented in Appendix~\ref{appendix_Wchi}.
The full matrix $\M{W}\M{\chi}^\frac12$ can be obtained from the Cholesky decomposition of the symmetric matrix $\M{W}\M{\chi}\V{W}$.
We note that if species $s$ is vanishing, then $(\M{W}\M{\chi}\M{W})_{is} = (\M{W}\M{\chi}\M{W})_{sj} = 0$, so no stochastic mass flux is generated for species $s$.

\begin{figure}
\centering
\setlength{\tabcolsep}{0pt} % for the horizontal padding
\begin{tabular}{|>{\centering\arraybackslash}m{2.5em}|>{\centering\arraybackslash}m{2.5em}|>{\centering\arraybackslash}m{2.5em}|>{\centering\arraybackslash}m{2.5em}|>{\centering\arraybackslash}m{2.5em}|>{\centering\arraybackslash}m{2.5em}|}
\hline
\cellcolor{yellow} \eqref{eq:Wchi_yellow} & \cellcolor{cyan} \eqref{eq:Wchi_cyan} & \cellcolor{yellow} & \cellcolor{cyan} & \cellcolor{yellow} & \cellcolor{yellow} \\
\hline
0 & \cellcolor{red} \eqref{eq:Wchi_red} & 0 & 0 & 0 & 0 \\
\hline
\cellcolor{yellow} & \cellcolor{cyan} & \cellcolor{yellow} & \cellcolor{cyan} & \cellcolor{yellow} & \cellcolor{yellow} \\
\hline
0 & 0 & 0 & \cellcolor{red} & 0 & 0 \\
\hline
\cellcolor{yellow} & \cellcolor{cyan} & \cellcolor{yellow} & \cellcolor{cyan} & \cellcolor{yellow} & \cellcolor{yellow} \\
\hline
\cellcolor{yellow} & \cellcolor{cyan} & \cellcolor{yellow} & \cellcolor{cyan} & \cellcolor{yellow} & \cellcolor{yellow} \\
\hline
\end{tabular}
\caption{Graphical depiction of the expansion of sub-matrix $\M{\chi}^\mathrm{sub}$ into a full matrix $\M{W}\M{\chi}$ for a 6-species system having vanishing species $w_2$ and $w_4$.  Depicted is the full matrix $\M{W}\M{\chi}$ where the colors correspond to the cases in \eqref{eq:Wchi}.}
\label{fig:expansion_matrix}
\end{figure}

We note that for each vanishing species only the diagonal element of $\M{W}\M{\chi}$ remains nonzero.
Hence, the diffusion of a dilute species $s$ ($w_s\ll 1$) becomes decoupled from other species (see \eqref{barF_N}) and the effective diffusion coefficient $D_s$ in \eqref{Deff} corresponds to the trace diffusion coefficient of $s$ in the given fluid mixture.
It is also important to note that the construction~\eqref{eq:Wchi} guarantees that $\overline{\V{F}}_s = \widetilde{\V{F}}_s = \V{0}$ for vanishing species, and ensures the mass conservation condition over all species, $\sum_{s'}\overline{\M{F}}_{s'} = \sum_{s'}\widetilde{\M{F}}_{s'} = \V{0}$.
Therefore, this procedure is robust to roundoff errors.

In our double-precision implementation, we treat any species $s$ with $w_s < 10^{-14}$ as a vanishing species.

\subsection{\label{subsec_temporal}Temporal Integration Scheme}

The spatial discretization of the non-reactive FHD equations for the mass densities yields a set of stochastic ordinary differential equations.
It is straightforward to incorporate our CME-based chemistry model from Section~\ref{subsec_chem} via additional Poisson-noise terms,
\begin{equation}
\label{PoissonSODE}
\rho_0 \frac{d w_{s,\V{i}}}{dt} = \Big[- \rho_0 \divg(w_s \V{v}) - \divg\V{F}_s\Big]_{\V{i}} 
+ \left[m_s \sum_r\sum_{\alpha=\pm} \D{\nu}^\alpha_{sr} \frac{\mathcal{P} (a_r^\alpha \D{V}dt)}{\D{V}dt}\right]_{\V{i}}.
\end{equation}
where $w_{s,\V{i}}$ denotes the mass fraction of species $s$ in cell $\V{i}$.

Our overall temporal integration strategy is a predictor-corrector approach for both species and velocity.
Our goal is to develop a scheme that is second-order accurate in space and time deterministically, exhibits second-order weak accuracy in time for the linearized FHD equations, and treats reactions in a manner consistent with the CME~\cite{KimNonakaBellGarciaDonev2017}.
As explained below in detail, we treat viscous momentum dissipation implicitly and species diffusion explicitly.
This is because in liquids the time step size is limited by the viscous Courant--Friedrichs--Lewy (CFL) condition (i.e., momentum diffusion) due to the large Schmidt number.

To combine a second-order midpoint tau-leap reaction sampling~\cite{AndersonMattingly2011, HuLiMin2011} with a predictor-corrector scheme for FHD, we adopt mass density updates from the ExMidTau (explicit midpoint tau leaping) scheme we previously developed for reaction-diffusion systems~\cite{KimNonakaBellGarciaDonev2017}.
Hence, our new scheme uses a \emph{midpoint} predictor for mass densities, which differs from our earlier \emph{trapezoidal} scheme for non-reactive FHD systems~\cite{DonevNonakaBhattacharjeeGarciaBell2015, NonakaSunBellDonev2015}.
We have compared several combinations of mass and momentum updates to identify the variant of the scheme that gives the most accurate spectrum of the fluctuations (structure factors) in both equilibrium and giant fluctuation settings. 
In Section~\ref{subsec_num_struct_fac}, we provide an analysis of the structure factors, which both guides and verifies the design decisions we made to ultimately choose this particular temporal integration scheme, and demonstrate the advantages of the midpoint scheme.
One important observation is that our temporal integrator is robust in the large Schmidt number limit, $\mathrm{Sc}=\nu/D\rightarrow\infty$, where $\nu = \eta/\rho_0$, unlike the trapezoidal scheme used in \cite{DonevNonakaBhattacharjeeGarciaBell2015, NonakaSunBellDonev2015}.

% Floating environments created with the float package (as algorithm is) are incompatible with the revtex4-1 class.
% The only way you have to include such an environment in your document, is to use the floating specifier H, which avoids the environment to float.
% If you want to let it float, you can insert the algorithm inside a figure environment without caption. In this way the algorithm caption also respects the revtex4-1 directives.
\begin{figure} 
\begin{algorithm}[H]
\caption{\label{alg:temporal}
Advancing the mass densities $\rho_s^n = \rho_0 w_s^n$ and velocity $\V{v}^n$ from time $t^n$ to $t^{n+1}=t^n+\D{t}$.
We list both centered and BDS options for advection; the BDS notation is defined in Section III.B. in \cite{NonakaSunBellDonev2015}.}
  \begin{enumerate}[nosep,leftmargin=*]
    \vspace{5pt}    
    \item Solve a predictor Stokes problem for the updated velocity $\V{v}^{n+1,*}$ and mechanical pressure $\pi^{n+\myhalf,*}$:
      \begin{subequations}
      \begin{align}
      \begin{split}
      \frac{\rho_0\V{v}^{n+1,*}-\rho_0\V{v}^n}{\D{t}} + \grad\pi^{n+\myhalf,*}
      &= -\divg\left(\rho_0\V{v}\V{v}^\mathrm{T}\right)^n + \frac12\divg\left( \eta^n \bar{\grad}\V{v}^n + \eta^n \bar{\grad}\V{v}^{n+1,*} \right) \\
      &+ \divg\left( \sqrt{\frac{\eta^n k_\mathrm{B}T}{\D{V}\D{t}} } \big(\overline{\M{\mathcal{W}}}^\text{mom}\big)^n \right) + \V{f}^n,
      \end{split} \\
      \divg\V{v}^{n+1,*} & = 0.
      \end{align}
      \end{subequations}
    \item Calculate predictor mass densities $\rho^{n+\myhalf,*}_s$ at the midpoint using the total diffusive mass fluxes $\V{F}^n$ as well as the reaction source term $R^n_s$ evaluated over the first half time step:
      \begin{align}
      \rho_s^{n+\myhalf,*} & = \textstyle \rho_s^n + \frac{\D{t}}{2} \left( -\divg\V{F}^n_s + m_s R^n_s \right) - \frac{\D{t}}{2} \divg
        \begin{cases}
        \rho_s^n \left( \frac{\V{v}^n+\V{v}^{n+1,*}}{2} \right) & \mbox{(centered)}, \\
        \textrm{BDS}\left(\rho_s^n,\frac{\V{v}^n+\V{v}^{n+1,*}}{2},\divg\V{F}^n_s,\frac{\Delta t}{2}\right) & \mbox{(BDS)},
        \end{cases}
        \\
      \divg\V{F}^n     & \equiv \textstyle \divg\left[ -\big(\rho_0\M{W}\V{\chi}\V{\Gamma}\grad\V{x}\big)^n + \sqrt{\frac{2\bar{m}\rho_0}{\D{V}\D{t}/2}} \big(\M{W}\M{\chi}^\frac{1}{2}\big)^n \Big(\M{\mathcal{W}}^\text{mass}_\paren{1}\Big)^n \right], \\
      \textstyle R^n_s & \equiv \textstyle \frac{1}{\D{V}\D{t}/2}\sum_r\sum_{\alpha=\pm} \D{\nu}^\alpha_{sr}\mathcal{P}_\paren{1}\big((a_r^\alpha)^n\D{V}\D{t}/2\big).
      \end{align}
    \item Calculate corrector mass densities $\rho^{n+1}_s$ at time $t^{n+1}$ using the total diffusive mass fluxes $\divg\V{F}^{n+\myhalf,*}$ as well as the reaction source term $R^{n+\myhalf,*}_s$ evaluated over the full time step:
      \begin{align}
      \begin{split}
      \rho_s^{n+1} & = \textstyle \rho_s^n + \D{t} \left( -\divg\V{F}^{n+\myhalf,*}_s + m_s R^{n+\myhalf,*}_s \right) \\
      & \quad - \D{t} \divg 
        \begin{cases}
        \rho_s^{n+\myhalf,*} \left( \frac{\V{v}^n + \V{v}^{n+1,*}}{2} \right) & \mbox{(centered)}, \\
        \textrm{BDS}\left(\rho_s^n,\frac{\V{v}^n+\V{v}^{n+1,*}}{2},\divg\V{F}^{n+\myhalf,*}_s,\Delta t\right) & \mbox{(BDS)},
        \end{cases}
      \end{split}
      \\
      \label{F2nd}
      \divg\V{F}^{n+\myhalf,*} & \equiv \textstyle \divg\left[ -\big(\rho_0\V{W}\V{\chi}\V{\Gamma}\grad\V{x}\big)^{n+\myhalf,*} 
      + \sqrt{ \frac{2\bar{m}\rho_0}{\D{V}\D{t}}} (\M{W}\M{\chi}^\frac{1}{2})^{n+\myhalf,*} \left(\frac{ \left(\M{\mathcal{W}}^\text{mass}_\paren{1}\right)^n + \left(\M{\mathcal{W}}^\text{mass}_\paren{2}\right)^n }{\sqrt{2}}\right)\right], \\
      R^{n+\myhalf,*}_s & \equiv \textstyle \frac12 \left[ R_s^n + \frac{1}{\D{V}\D{t}/2} \sum_r\sum_{\alpha=\pm} \D{\nu}^\alpha_{sr} \mathcal{P}_\paren{2}\left( \left( 2 (a_r^\alpha)^{n+\myhalf,*} - (a_r^\alpha)^n \right)^+ \D{V}\D{t}/2 \right) \right].
      \end{align}
    \item Solve a corrector Stokes problem for the updated velocity $\V{v}^{n+1}$ and mechanical pressure $\pi^{n+\myhalf}$:
      \begin{subequations}
      \begin{align}
      \begin{split}
      \frac{\rho_0\V{v}^{n+1}-\rho_0\V{v}^n}{\D{t}} + \grad\pi^{n+\myhalf}
      &= -\frac12\divg\left[\left(\rho_0\V{v}\V{v}^\mathrm{T}\right)^n + \left(\rho_0\V{v}\V{v}^\mathrm{T}\right)^{n+1,*}\right] 
      + \frac12\divg\left( \eta^n \bar{\grad}\V{v}^n + \eta^{n+1} \bar{\grad}\V{v}^{n+1} \right) \\
      &+ \frac12\divg\left[ \left( \sqrt{\frac{\eta^n k_\mathrm{B}T}{\D{V}\D{t}} } + \sqrt{\frac{\eta^{n+1} k_\mathrm{B}T}{\D{V}\D{t}} } \right) \big(\overline{\M{\mathcal{W}}}^\text{mom}\big)^n \right] + \V{f}^{n+\myhalf,*}, 
      \end{split} \\
      \divg\V{v}^{n+1} & = 0.
      \end{align}
      \end{subequations}
  \end{enumerate}
\end{algorithm}
\end{figure}

We advance the system from time $t_n = n\D{t}$ to time $t_{n+1} = (n+1)\D{t}$ in four steps:
\begin{enumerate}
\item Perform a predictor Stokes solve for the velocity $\V{v}^{n+1,*}$ at $t_{n+1}$.
\item Calculate predictor mass densities $\rho^{n+\myhalf,*}_s$ at the midpoint time $t=t_n+\frac12\D{t}$.
\item Calculate corrector mass densities $\rho^{n+1}_s$ at time $t_{n+1}$.
\item Perform a corrector Stokes solve for velocity $\V{v}^{n+1}$ at $t_{n+1}$.
\end{enumerate}
These steps are elaborated in detail in Algorithm~\ref{alg:temporal}.
In the algorithm description, superscripts are used to denote a time level where a given quantity is evaluated, e.g., $\V{f}^n=\V{f}(\V{w}^n)$.
Also, $\big(\M{\mathcal{W}}^\text{mom}\big)^n$ and $\Big(\M{\mathcal{W}}^\text{mass}_\paren{i}\Big)^n$ ($i=1,2$) denote collections of i.i.d.\ (independent and identically distributed) standard normal random variables generated on control volume faces independently at each time step, and $\overline{\M{\mathcal{W}}}^\text{mom} \equiv \M{\mathcal{W}}^\text{mom} + \big(\M{\mathcal{W}}^\text{mom}\big)^\mathrm{T}$.
We denote collections of independent Poisson random variables generated at cell centers independently at each time step with $\mathcal{P}_\paren{i}$ ($i=1,2$), and denote $[\bullet]^+ \equiv \max(\bullet,0)$.

In our time-advancement scheme, each Stokes problem couples a Crank--Nicolson discretization of viscous dissipation to the divergence-free constraint on velocity, to simultaneously solve for the velocity and mechanical pressure.
To solve the Stokes system we use a variable-coefficient (tensor) multigrid-preconditioned GMRES (generalized minimal residual) solver~\cite{CaiNonakaBellGriffithDonev2014}, as we have done previously~\cite{NonakaSunBellDonev2015, DonevNonakaBhattacharjeeGarciaBell2015}.
The difference between the predictor and corrector Stokes solves is the temporal discretization of the advective term (explicit vs.\ trapezoidal) and the forcing term (explicit vs.\ midpoint); both Stokes solves are required for second-order deterministic accuracy.

As mentioned above, Steps~2 and 3 of the present scheme become essentially the same as the ExMidTau scheme in the dilute limit in the absence of advection.
The only difference is a Stratonovich-type update of the stochastic mass flux in \eqref{F2nd}.
While our previous analysis for RDME systems~\cite{KimNonakaBellGarciaDonev2017} adopted the Ito interpretation, we choose the Stratonovich-type update here since a general analysis for weak fluctuations (linearized FHD)~\cite{DelongGriffithVandenEijndenDonev2013} guarantees second-order weak accuracy of the overall scheme for this choice.
It can be shown that the Stratonovich and Ito interpretations become identical in the dilute limit.
Hence, our numerical method not only achieves second-order weak accuracy for weak fluctuations but also inherits nice features of the ExMidTau scheme carefully designed for strong fluctuations.

\subsection{\label{subsec_num_struct_fac}Structure Factor Analysis}

We analyze our new temporal integrator by investigating time integration errors in the spectrum of giant concentration fluctuations for a binary mixture undergoing a dimerization reaction.
We assume that a weak uniform concentration gradient is applied along the $y$-axis with gravity pointing in the positive $y$-direction.
The Fourier-transformed linearized equations for $\d{v}_\parallel\equiv \d{v}_y$ and $\d{w}$ (see Appendix~C in \cite{BhattacharjeeBalakrishnanGarciaBellDonev2015}) take the form:
\begin{subequations}
\label{linGF}
\begin{align}
\partial_t (\d{\hat{v}}_\parallel) &= -\nu k^2 (\d{\hat{v}}_\parallel) \! + \! \sqrt{2 \nu k_\mathrm{B}T\rho_0^{-1}}\: i\V{k} \! \cdot \! \hat{\V{\mathcal{Z}}}^\mathrm{mom} \! + \! g \zeta (\d{\hat{w}}),\\
\partial_t (\d{\hat{w}}) &= -h (\d{\hat{v}}_\parallel) \! - \! Dk^2 (\d{\hat{w}}) \! + \! \sqrt{2 D k_\mathrm{B}T \rho_0^{-1}\mu_w^{-1}}\: i\V{k} \! \cdot \! \hat{\V{\mathcal{Z}}}^\mathrm{mass} \! - \! r (\d{\hat{w}}) \! + \! \sqrt{2 r k_\mathrm{B}T \rho_0^{-1}\mu_w^{-1}}\: \hat{\mathcal{Z}}^\mathrm{react}.
\end{align}
\end{subequations}
Here $\V{k}\equiv\V{k}_\perp$ is a wavevector in the plane perpendicular to the gradient, $g$ is the gravitational acceleration, $\zeta=\rho^{-1}(\partial\rho/\partial w)$ is the solutal expansion coefficient, and $h$ is the concentration gradient, $\grad w = h\V{e}_y$.
Using the method developed in \cite{DonevVandenEijndenGarciaBell2010}, we analytically compute the resulting structure factors when our temporal integrator is used to solve \eqref{linGF}.
For the non-reactive case ($r=0$), we also compute structure factors obtained from two schemes developed in our previous work~\cite{DonevNonakaBhattacharjeeGarciaBell2015, NonakaSunBellDonev2015}.
The overdamped scheme (see Algorithm~2 in \cite{NonakaSunBellDonev2015}) uses the steady Stokes equation, i.e., eliminates the inertial term $\partial_t \V{v}=\V{0}$ by taking an overdamped limit.
We refer to the previous scheme for solving the inertial equation as the inertial trapezoidal scheme (see Algorithm~1 in \cite{NonakaSunBellDonev2015}), and to our new scheme as the inertial midpoint scheme (see Algorithm~\ref{alg:temporal}).

We set $\D{x}=1$ and $k_\mathrm{B}T/\rho_0 = 1$.
To denote how fast momentum diffusion, species diffusion, and reaction are, we define the following dimensionless Courant numbers:
\begin{equation}
\label{unitlessvars}
\alpha=\frac{\nu\D{t}}{\D{x}^2},\quad
\beta=\frac{D\D{t}}{\D{x}^2},\quad
\gamma=r\D{t}.
\end{equation}
To consider the case of a relatively large $\D{t}$ with a large Schmidt number $\mathrm{Sc}=10^3$ (as is typical of liquid mixtures), we set $\alpha=250$ and $\beta=0.25$.
For the reactive case, we consider two reaction rates $\gamma=0.025$ and $\gamma=0.1$, corresponding to penetration depths $\xi=\sqrt{D/r}=\sqrt{10}\D{x}$ and $\xi=\frac12\sqrt{10}\D{x}$, respectively.
Other parameters are chosen so that $\mu_w h^2\D{t}^2=100$, and $g\zeta h\D{t}^2=0.025$ if gravity is present.

\begin{figure}
\centerfloat
\includegraphics[width=1.2\textwidth]{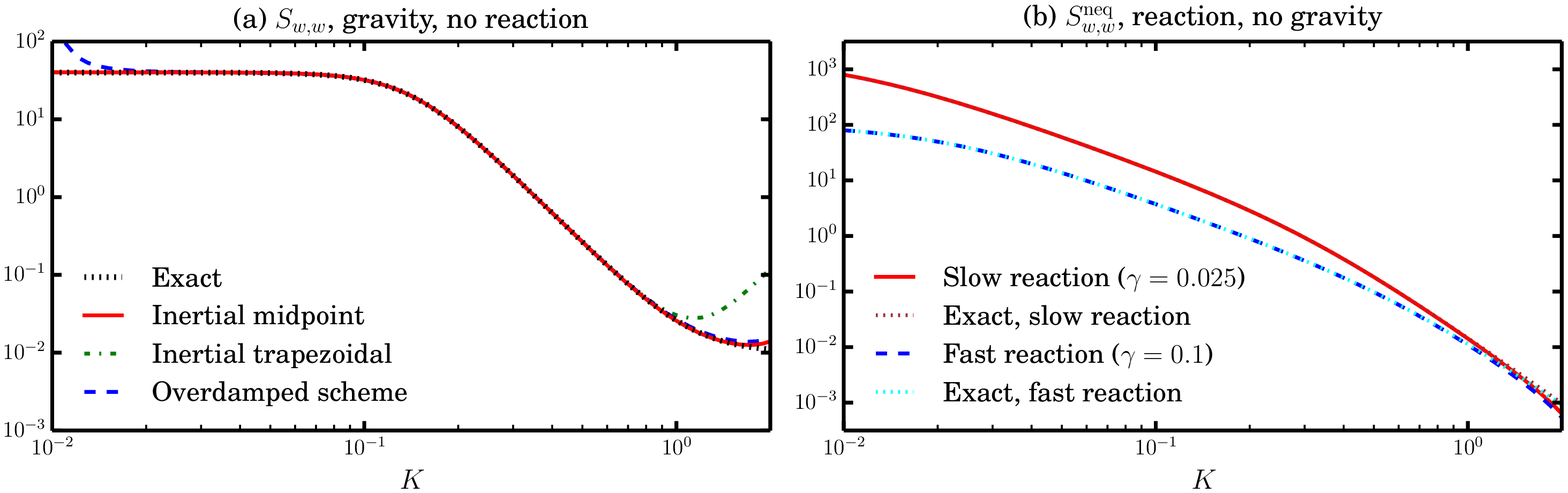}
\caption{\label{fig_struct_fac_analysis}
Structure factors for giant concentration fluctuations.
Panel~(a) shows $S_{w,w}$ for the non-reactive case with gravity.
Results from our numerical scheme (inertial midpoint) and two earlier schemes are compared with the exact result.
Panel~(b) shows $S_{w,w}^\mathrm{neq}=S_{w,w}-S_{w,w}^\mathrm{eq}$ for the reactive case with no gravity.
For two rate constants, results from our numerical scheme are compared with the exact result.
The $x$-axis is the dimensionless wavenumber $K=k\D{x}$.}
\end{figure}

The structure factor can be decomposed into the sum $S_{w,w}=S^\mathrm{eq}_{w,w} + S^\mathrm{neq}_{w,w}$, where $S^\mathrm{eq}_{w,w}$ is the equilibrium structure factor \eqref{Seq} and $S^\mathrm{neq}_{w,w}$ is the nonequilibrium enhancement.
In the non-reactive case with no gravity, the nonequilibrium enhancement exhibits a $k^{-4}$ power law in the entire range of wavenumbers $k$,
\begin{equation}
S^\mathrm{neq}_{w,w} = \frac{k_\mathrm{B}T}{\rho_0 D(D+\nu)k^4}h^2.
\end{equation}
However, the power law is suppressed at small $k$ by gravity~\cite{DonevNonakaSunFaiGarciaBell2014} or reaction~\cite{BhattacharjeeBalakrishnanGarciaBellDonev2015}.

For the non-reactive case with gravity, we compare $S_{w,w}$ obtained from the three schemes with the exact result in Fig.~\ref{fig_struct_fac_analysis}~(a). 
A power-law spectrum $S_{w,w}\sim k^{-4}$ develops for intermediate wavenumbers $k$.
At small wavenumbers, $S_{w,w}$ becomes constant due to gravity.
At large wavenumbers, the $k^{-4}$ decay in the nonequilibrium part is hidden due to the flat equilibrium structure factor $S^\mathrm{eq}_{w,w}$.
Our numerical scheme reproduces $S_{w,w}$ accurately for all but the largest $k$ values, whereas both earlier schemes exhibit significant deviations at either large or small $k$ values.
Significant deviations of the previous inertial scheme at large $k$ are due to temporal integration errors in the nonequilibrium part $S^\mathrm{neq}_{w,w}$, as can be seen more clearly by examining the cross-correlation between fluctuations of $w$ and $v$ (not shown).
The divergence of $S_{w,w}$ for the overdamped scheme at small $k$ demonstrates that the overdamped limit does not apply for sufficiently small $k$ with gravity.
Thus, our new scheme combines the favorable features of our previous trapezoidal inertial scheme (correct behavior for small $k$ with gravity) and the overdamped scheme (correct behavior for large $k$).

In Fig.~\ref{fig_struct_fac_analysis}~(b), we show the nonequilibrium enhancement $S^\mathrm{neq}_{w,w}$ in the structure factor for the reactive case with no gravity.
We obtain the exact $S^\mathrm{neq}_{w,w}$ by analyzing \eqref{linGF} without the stochastic mass fluxes and with deterministic reaction (see also Eq. (58) in~\cite{ZarateSengersBedeauxKjelstrup2007} or Eq. (44) in~\cite{BedeauxZaratePagonabarragaSengersKjelstrup2011}),
\begin{equation}
\label{Sneqww}
S^\mathrm{neq}_{w,w}=\frac{k_\mathrm{B}T}{\rho_0(Dk^2+r)[(D+\nu)k^2+r]} h^2.
\end{equation}
Our midpoint scheme reproduces $S^\mathrm{neq}_{w,w}$ accurately for both rate constants.
We emphasize that these results are remarkable given that $\alpha=\mathrm{O}(10^2)$.
Our new scheme remains accurate for $\alpha=\beta\mathrm{Sc}\gg 1$ because the relative error in $S^\mathrm{neq}_{w,w}$ for our midpoint scheme has the form $[1+\mathcal{O}(\mathrm{Sc}^{-1})]\mathcal{O}(\D{t}^2)$, indicating robust behavior for large Schmidt numbers.
On the other hand, the relative error for the trapezoidal scheme has an $\mathcal{O}(\mathrm{Sc})\mathcal{O}(\D{t}^4)$ term, which results in significant deviations at large $k$ as observed in Fig.~\ref{fig_struct_fac_analysis}~(a).

\section{\label{sec_examples}Numerical Examples}

In this section, we consider four examples that demonstrate the capabilities of our numerical methodology.
In Section~\ref{subsec_sucrose}, we model the hydrolysis of sucrose in an aqueous solution with very dilute solutes.
In Section~\ref{subsec_dimer_dist}, we investigate a binary liquid mixture undergoing a dimerization reaction at thermodynamic equilibrium.
In Section~\ref{subsec_dimer_giant_fluct}, to verify the correct coupling of mass and momentum fluctuations, we study nonequilibrium giant fluctuations in a mixture undergoing a dimerization reaction.
In Section~\ref{subsec_fingering}, to demonstrate the scalability and practical utility of our method, we investigate the effects of fluctuations for a reactive fingering instability.

\subsection{\label{subsec_sucrose}Hydrolysis of Sucrose}

We consider a dilute solution of sugar in water at equilibrium, undergoing the reversible hydrolysis reaction
\begin{equation}
\mathrm{sucrose} + \mathrm{H}_2\mathrm{O} \rightleftharpoons \mathrm{glucose} + \mathrm{fructose}.
\end{equation}
Sucrose is particularly dilute, with only $\sim 10$ molecules per computational cell, whereas there are $\sim 10^7$ glucose and fructose molecules and $\sim 10^{10}$ water molecules per cell.
We investigate the equilibrium distribution of the number of sucrose molecules in a cell to demonstrate that our approach correctly models the dilute limit.

We use cgs units and choose physical parameters assuming $T=293$, atmospheric pressure, $\rho_0 = 1$, and $\eta = 0.01$.
The four species are glucose ($s=1$), fructose ($s=2$), sucrose ($s=3$), and water ($s=4$).
Using the trace diffusion coefficients of the solutes, $D_s$ ($s=1,2,3$)~\cite{VenancioTeixeira1997, TilleyMills1967}, and the self-diffusion coefficient of water $D_\mathrm{water}$~\cite{KrynickiGreenSawyer1978}, the Maxwell--Stefan binary diffusion coefficients are assigned as in \cite{LiuBardowVlugt2011},
\begin{equation}
\label{MSdiffcoeff}
\textsc{\DJ}_{s 4} = D_s,\quad \textsc{\DJ}_{s s'}=\frac{D_s D_{s'}}{D_\mathrm{water}}\quad(s,s'=1,2,3).
\end{equation}
Since we consider the dilute limit, we assume that the system is an ideal mixture and obeys the traditional LMA, with forward rate $a^+=k^+ n_3$, reverse rate $a^-=k^-n_1 n_2$, and equilibrium constant $K=n_1^\mathrm{eq} n_2^\mathrm{eq} / n_3^\mathrm{eq}$.
The reaction equilibrium lies almost completely in the direction of the formation of glucose and fructose~\cite{GoldbergTewariAhluwalia1989}, but uncatalyzed sucrose hydrolysis is extremely slow (with a half-life of 500 years)~\cite{WolfendenYuan2008}.
While we use an experimental value of $K$~\cite{GoldbergTewariAhluwalia1989}, we artificially increase the reaction rates to $k^+ = 10$ and $k^- = K/k^+$ so that the forward reaction occurs about 100 times per cell per simulation.

We set up a two-dimensional system consisting of $32\times 32$ cells with dimensions $\D{x}=\D{y}=10^{-4}$ and periodic boundary conditions.
The thickness of the system is $\D{z}=10^{-4}$ and the cell volume $\D{V}=\D{x}\D{y}\D{z}$.
We consider the case where there are ten sucrose molecules per cell.
Hence, $n_3^\mathrm{eq}$ is determined from $n_3^\mathrm{eq}\D{V} = 10$ and $n_1^\mathrm{eq}=n_2^\mathrm{eq}$ are subsequently determined from equilibrium.
The resulting equilibrium mass fractions are $w_1^\mathrm{eq} = w_2^\mathrm{eq} = \num{4.9e-03}$, $w_3^\mathrm{eq} = \num{5.7e-09}$, and $w_4^\mathrm{eq} = 0.990$.
We use two time step sizes, $\D{t}=10^{-5}$ and $10^{-4}$, to check the continuous-time limit and quantify time integration errors.
Note that the larger $\D{t}$ corresponds to diffusive CFL numbers $D_{s,\rm max}\D{t}/\D{x}^2=0.07$ for species diffusion and $\nu\D{t}/\D{x}^2 = 100$ for momentum diffusion.
For each value of $\D{t}$, we ran 16 independent samples up to time $\mathcal{T}=1$, collecting data every $t=10^{-4}$ for $t\ge\mathcal{T}/10$.

\begin{figure}
\centerfloat
\includegraphics[width=0.75\textwidth]{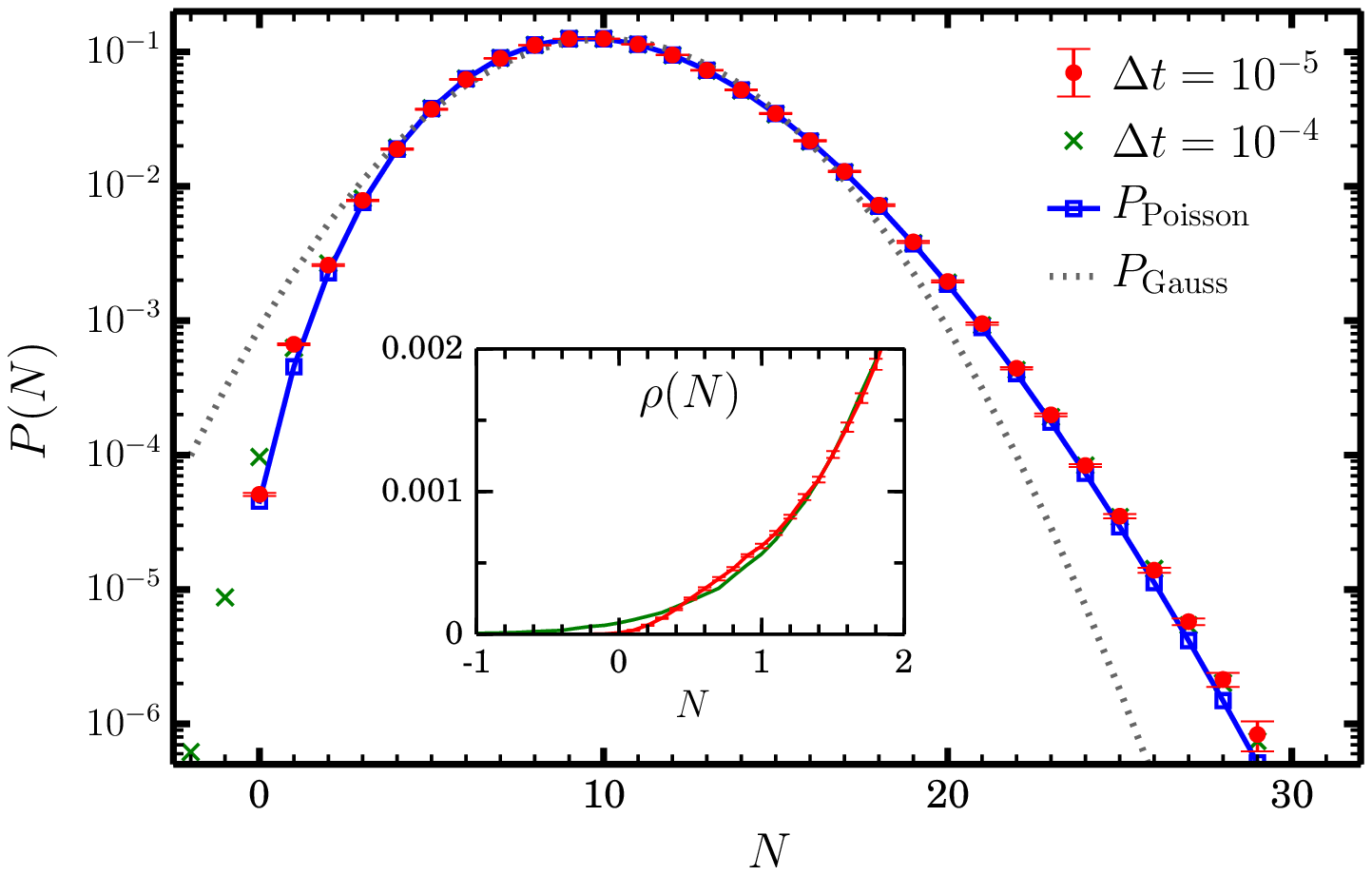}
\caption{\label{fig_sucrose}
Equilibrium distribution for a dilute sugar solution undergoing a hydrolysis reaction.
Numerical results for the distribution (histogram) $P(N)$ of the number of sucrose molecules in a cell $N$ are compared with the physically correct Poisson distribution $P_\mathrm{Poisson}(N)$, and its Gaussian approximation $P_\mathrm{Gauss}(N)$.
In the inset, numerical results for the continuous distribution $\rho(N)$ are shown near $N=0$.
Results from a smaller time step size $\D{t}=10^{-5}$ are plotted with error bars corresponding to two standard deviations, whereas those from a larger time step size $\D{t}=10^{-4}$ are plotted without errorbars for clarity.}
\end{figure}

We recall that the number of sucrose molecules in a cell $N=n_3\D{V}$ has a continuous range in FHD simulations.
We define its discrete distribution as $P(N) = \int_{N-\frac12}^{N+\frac12}\rho(N') dN'$, where $\rho(N)$ is the continuous distribution of $N$.
Figure~\ref{fig_sucrose} shows that for the smaller $\D{t}$, $P(N)$ is remarkably close to the physically correct Poisson distribution $P_\mathrm{Poisson}(N)$, and $\rho(N)$ is essentially zero for negative values of $N$.
We note that $P_\mathrm{Poisson}(N)$ is significantly different from its Gaussian approximation $P_\mathrm{Gauss}(N)$.
For the larger $\D{t}$, while the remarkable agreement with the Poisson distribution is still observed, negative values of $N$ start to appear, yielding $\int_{-\infty}^0\rho(N)dN \approx \num{3e-5}$, see the inset in Fig.~\ref{fig_sucrose}.
The same results were obtained in our previous reaction-diffusion model of dilute solutions~\cite{KimNonakaBellGarciaDonev2017}, confirming that our treatment of the stochastic mass flux coefficients (see Section~\ref{subsec_spatial}) is consistent with the dilute limit, even in the presence of random advection.
We have also confirmed that the equilibrium structure factor of each species (not shown) has a flat spectrum (as predicted by theory~\cite{DonevNonakaBhattacharjeeGarciaBell2015}), indicating that there are no spurious correlations between cells.

\subsection{\label{subsec_dimer_dist}Dimerization: Equilibrium Distribution}

We next consider a liquid mixture undergoing the dimerization reaction~\eqref{dimer_rxn}.
This binary system contains monomers $\mathrm{A}$ ($s=1$) and dimers $\mathrm{A}_2$ ($s=2$) and is representative of cyclic dimer formation in pure liquid acetic acid.
We demonstrate here our ability to model a system with strong fluctuations in the absence of a dominant solvent by considering a small number of molecules ($\sim 10$) of each species per cell.
As in the sugar solution example, we investigate the equilibrium distribution of monomers and dimers; however, since the system is not dilute, the distribution of each species is not Poisson.
The numbers of monomers and dimers in a cell ($N_1$ and $N_2$), do not vary independently due to the constant density assumption $N_1 + 2 N_2 = \rho_0\D{V}/m$, where $m$ is the mass of a monomer.
Therefore, we investigate the equilibrium distribution of $N_2$, $P(N_2)$, for $N_1+2N_2=40$.

We simulate a two-dimensional system consisting of $32 \times 32$ cells under periodic boundary conditions.
Here we use arbitrary units that give $\D{x}=\D{y}=\D{z}=1$, $\textsc{\DJ}_{12}=1$, and $m = 1$ with $k_\mathrm{B}=1$.
We set $\rho_0=40$ with $w_1^\mathrm{eq} = w_2^\mathrm{eq} = 0.5$ so that $N_1 + 2N_2=40$ with $N_1^\mathrm{eq}=20$ and $N_2^\mathrm{eq}=10$.
We set the reaction rates $a^\pm$ as in \eqref{corrLMA} with a modification 
\begin{equation}
\label{corrLMA_integer}
a^+ = \kappa^+ \left(\frac{N_1^+}{N_1^+ + N_2^+}\right)\left(\frac{(N_1-1)^+}{N_1^+ +N_2^+ -1}\right),
\end{equation}
where $N^+ =\max(N,0)$.
Note that \eqref{corrLMA_integer} turns off unphysical reactions when $0<N_1<1$.
The rate constants $\kappa^+=0.8724$ and $\kappa^-=1.125$ are chosen as follows.
The ratio $K=\kappa^+/\kappa^-= 0.7755$ is determined so that the resulting theoretical distribution gives $\langle N_2\rangle = \sum_{N_2}N_2 P(N_2) = N_2^\mathrm{eq}$.
The magnitude of $\kappa^\pm$ is determined so that the linearized reaction rate $r=0.1$ (see \eqref{dimer_lin_rate}) gives a penetration depth $\xi\equiv\sqrt{\textsc{\DJ}_{12}/r}=\sqrt{10}\D{x}$.
We set $\eta=10^3$ and $\mathcal{T}=10^3$.
We use a small $\D{t}=10^{-2}$ to minimize temporal integration errors.
For 16 independent samples with $10^5$ time steps, we collect data every $10^2$ time steps, discarding the first $10^4$ time steps.

\begin{figure}
\centerfloat
\includegraphics[width=0.75\textwidth]{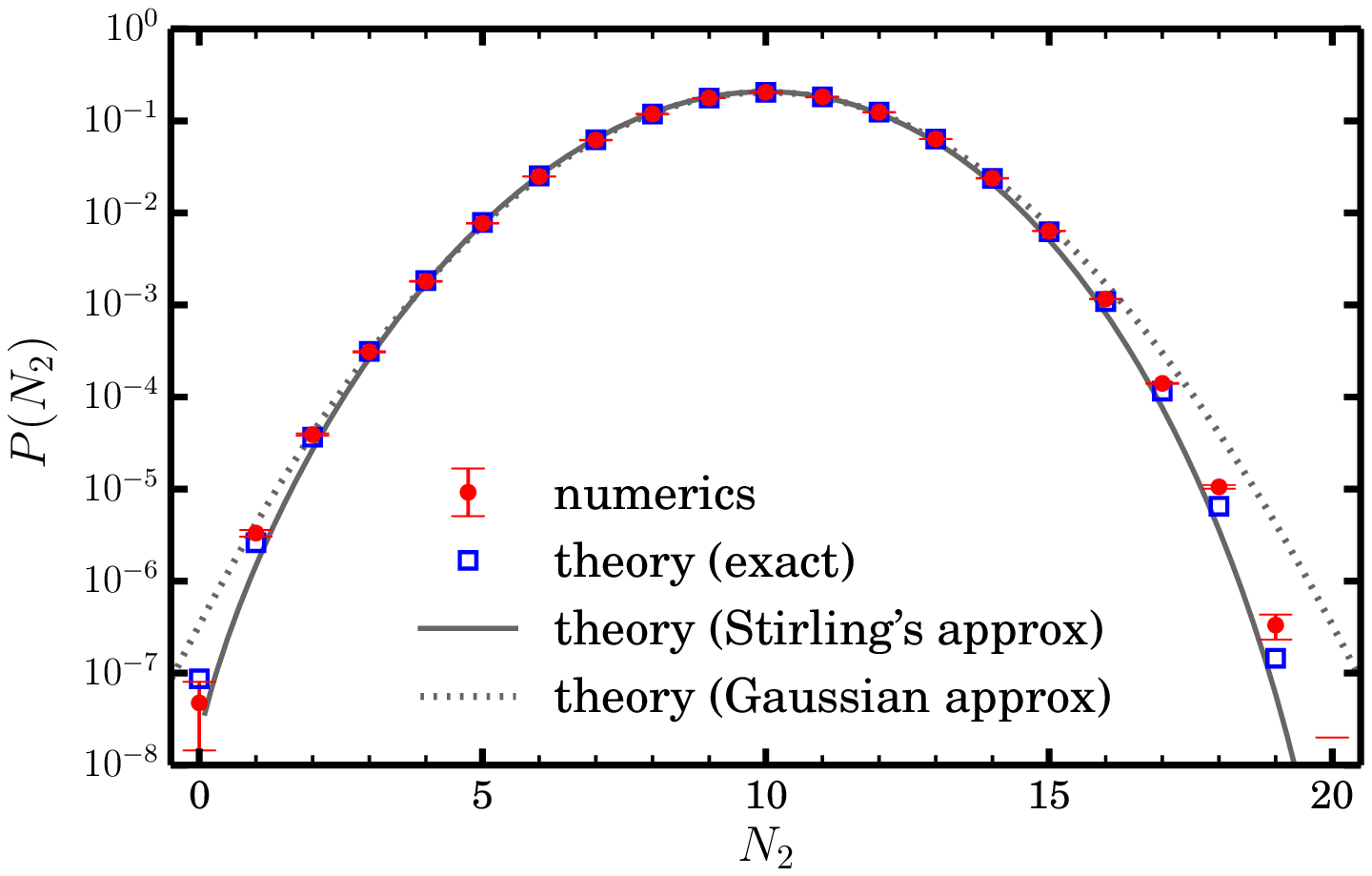}
\caption{\label{fig_dimer_hist}
Equilibrium distribution for a binary ideal mixture undergoing a dimerization reaction.
The distribution $P(N_2)$ of the number of dimers in each cell $N_2$ is computed using our numerical method and compared with various theoretical results (see text).
Error bars correspond to two standard deviations.}
\end{figure}

In Fig.~\ref{fig_dimer_hist}, we compare the simulation result for the equilibrium distribution $P(N_2)$ with theoretical results obtained in Section~\ref{subsubsec_formul_dimer_eq_dist}.
We denote the exact Einstein distribution obtained from the entropy expression~\eqref{SN1N2} by $P_\mathrm{exact}$, the Stirling's approximation result obtained from \eqref{S_Stirling} by $P_\mathrm{Stirling}\sim\exp(S_\mathrm{Stirling}/k_\mathrm{B})$, and the Gaussian approximation of $P_\mathrm{Stirling}$ by $P_\mathrm{Gauss}$.
Note that $P_\mathrm{exact}$ is a discrete distribution whereas $P_\mathrm{Stirling}$ and $P_\mathrm{Gauss}$ have continuous ranges, $0<N_2<20$ and $-\infty<N_2<\infty$, respectively.
Significant deviations of $P_\mathrm{Gauss}$ from $P_\mathrm{exact}$ indicate that the system is subject to strong fluctuations, as expected from the small number $N_2^\mathrm{eq}=10$.
Over a remarkably wide range, our numerical method accurately matches $P_\mathrm{exact}$.
Even beyond this range, it gives sensible values with accuracy better than or comparable to $P_\mathrm{Stirling}$.
Measurable deviations are observed only for larger values $N_2=19$ and 20, for which the occupation probabilities are already very small ($P_\mathrm{exact}(N_2)<10^{-6}$).

It is important to note that statistically identical results for $P(N_2)$ are obtained from the corresponding non-reactive system with $\kappa^\pm=0$ (not shown).
This confirms thermodynamic consistency of our overall formulation. 
In addition, this also confirms the validity of our overall numerical treatment for diffusion with strong fluctuations. 
In particular, considering that our multiplicative GWN modeling for strong fluctuations was developed in the dilute limit~\cite{KimNonakaBellGarciaDonev2017} and analyzed only for this case, the validity of its extension to non-dilute solutions cannot be taken for granted.

\subsection{\label{subsec_dimer_giant_fluct}Dimerization: Giant Fluctuations}

We now investigate a system where velocity fluctuations are coupled to diffusion.
We consider the same dimerization reaction, but examine giant fluctuations in the presence of a weak concentration gradient with no gravity.
We focus on the nonequilibrium contribution to the structure factor, $S^\mathrm{neq}_{w_1,w_1}=S_{w_1,w_1}-S^\mathrm{eq}_{w_1,w_1}$, for wavevectors perpendicular to the concentration gradient.
We neglect stochastic mass fluxes and use deterministic chemistry so that we eliminate the equilibrium fluctuations, and thus obtain $S^\mathrm{neq}_{w_1,w_1}$ with greater statistical accuracy.
We have previously considered a gas mixture in a similar setting~\cite{BhattacharjeeBalakrishnanGarciaBellDonev2015}; here we consider a liquid mixture with a large Schmidt number $\mathrm{Sc}=10^3$, which quantitatively changes the spectrum of giant fluctuations.

A detailed theoretical analysis of giant fluctuations using linearized FHD first appeared in \cite{ZarateSengersBedeauxKjelstrup2007} assuming that the system is near chemical equilibrium everywhere.
It was later extended in \cite{BedeauxZaratePagonabarragaSengersKjelstrup2011} to account for the nonlinearity caused by the fact that the system is not everywhere in chemical equilibrium; this theoretical analysis assumes a liquid mixture so it was not applicable for the gas mixture we considered in \cite{BhattacharjeeBalakrishnanGarciaBellDonev2015}.
In these theoretical studies the concentration gradient was imposed via the Soret effect by applying a temperature gradient, unlike the case we consider here where the concentration is fixed at the $y$-walls using reservoir boundary conditions. 
Furthermore, the theoretical studies in \cite{ZarateSengersBedeauxKjelstrup2007, BhattacharjeeBalakrishnanGarciaBellDonev2015} did not account for the boundary conditions for the fluctuating fields.

We consider a two-dimensional square domain of side length $L_x = L_y =64$ (in arbitrary units), and periodic boundary conditions in the $x$-direction.
The system is divided into $128 \times 128$ grid cells with grid spacing $\D{x}=\D{y}=0.5$.
To remain in the linearized FHD regime, we increase the cell depth to $\D{z}=10^6$ so that there are sufficiently many monomers and dimers in a cell, $N_1+2N_2=\num{2.5e5}$ for $\rho_0 = 1$ and $m=1$. 
We set $\textsc{\DJ}_{12}=1$, $\eta=10^3$, and $k_\mathrm{B}T=10^3$.
For the dimerization reaction, the equilibrium constant $K = \kappa^+/\kappa^-=0.75$ is chosen to give a reference equilibrium state with $w_1^\mathrm{eq}=w_2^\mathrm{eq}=0.5$.
Two sets of reaction constants are considered: $(\kappa^+,\kappa^-)=(\num{8.438e-2},0.1125)$, corresponding to linearized reaction rate $r=0.4$ and penetration depth $\xi=\sqrt{10}\D{x}$; and $(\kappa^+,\kappa^-)=(\num{8.438e-3},\num{1.125e-2})$, corresponding to $r=0.04$ and $\xi=10\D{x}$.
The time step size is set to $\D{t}=0.025$, which gives Courant numbers $\textsc{\DJ}_{12}\D{t}/\D{x}^2=0.1$ and $\nu\D{t}/\D{x}^2=100$.
We ran $10^5$ steps discarding the first $10^4$ steps, and computed the steady-state monomer concentration profile $\bar{w}_1(y)$ and $S^\mathrm{neq}_{w_1,w_1}(k_x)$.

\begin{figure}
\centerfloat
\includegraphics[width=1.2\textwidth]{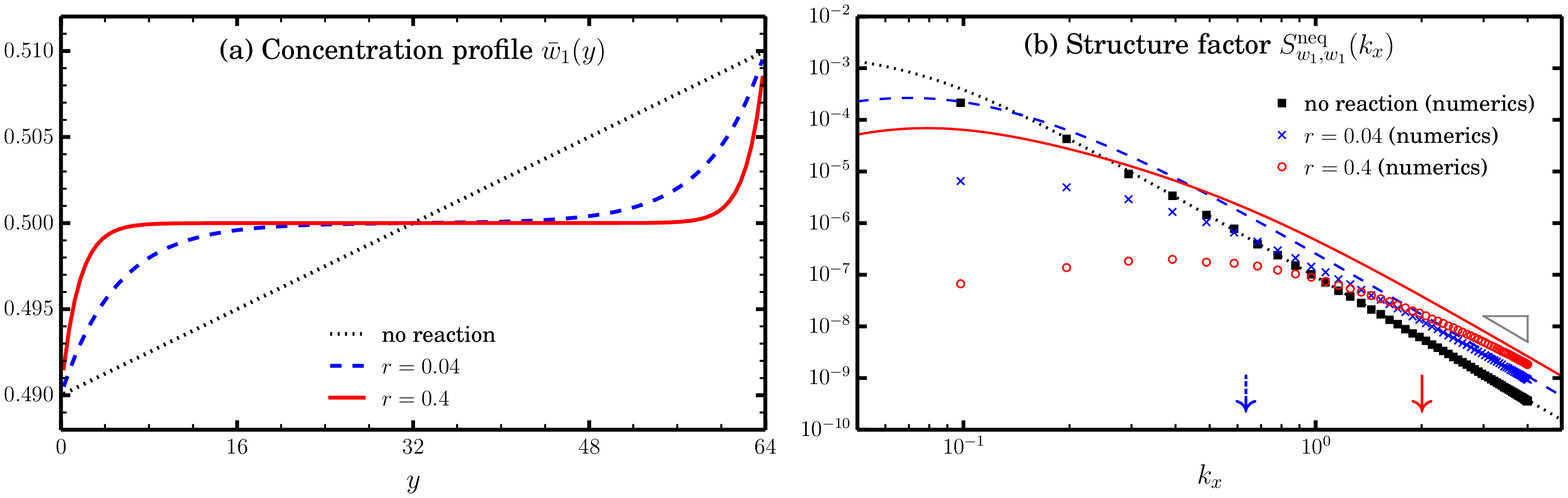}
\caption{\label{fig_dimer_giant_fluct}
Giant fluctuations with a dimerization reaction.
Panel~(a) Steady-state monomer concentration profile $\bar{w}_1(y)$.
Results from two linearized reaction rates $r$ and the non-reactive case are compared.
Panel~(b) Nonequilibrium enhancement $S^\mathrm{neq}_{w_1,w_1}$ in the structure factor of the monomer mass fraction.
Numerical results (depicted by symbols) are compared with theoretical predictions obtained under a linear gradient setting (depicted by a solid line for $r=0.4$, dashed for $r=0.04$, and dotted for $r=0$).
A slope marker for the $k^{-4}$ decay is drawn and arrows denoting $k_x = \sqrt{10r/\textsc{\DJ}_{12}}$ are depicted for $r=0.4$ (solid) and $r=0.04$ (dashed).
Note that nonlinear gradients develop in reactive cases, which explains discrepancies between numerical and theoretical results at small to intermediate wavenumbers.}
\end{figure}

To impose a concentration gradient in the $y$-direction, no-slip reservoir conditions~\cite{DonevNonakaSunFaiGarciaBell2014} are imposed with $(w_1,w_2)=(0.49,0.51)$ at $y=0$ and $(w_1,w_2)=(0.51,0.49)$ at $y=L_y$.
While a linear concentration profile is formed in the steady state for the non-reactive case, a nonlinear profile is generated by the dimerization reaction.
In Fig.~\ref{fig_dimer_giant_fluct}~(a), the profiles of $\bar{w}_1 (y)$ for reaction rates $r=0.4$ and 0.04 are compared with the one for the non-reactive case.
As $r$ increases, the nonlinearity in $\bar{w}_1(y)$ becomes more evident. 
This is because a larger region around $y=L_y/2$ is constrained to be in chemical equilibrium due to faster reactions, resulting in larger concentration gradients at the boundaries.
Identical concentration profiles are obtained from the corresponding deterministic reaction-diffusion systems (not shown).

In Fig.~\ref{fig_dimer_giant_fluct}~(b), we show numerical results of $S^\mathrm{neq}_{w_1,w_1}$.
To account for errors in the discrete approximation to the continuum Laplacian, the modified wavenumber~\cite{UsabiagaBellDelgadoBuscalioniDonevFaiGriffithPeskin2012}
\begin{equation}
\tilde{k}_x = \frac{\sin(k_x\D{x}/2)}{\D{x}/2}
\end{equation}
is used instead of $k_x$.
For the non-reactive case ($r=0$), a clear $k^{-4}$ power law is observed until the confinement effect becomes significant for small $k_x \ll L_y^{-1}$.
For the reactive cases, the $k^{-4}$ power law is only observed at large $k_x \gg \sqrt{r/\textsc{\DJ}_{12}}$.
For larger $r$, the $k^{-4}$ power law appears in a narrower range of $k_x$ values and the prefactor of the power law becomes larger.

For the non-reactive case, the prefactor of the power law is accurately predicted by the theoretical prediction~\eqref{Sneqww}.
By multiplying \eqref{Sneqww} by the confinement factor~\cite{ZarateKirkpatrickSengers2015}
\begin{equation}
\label{confinefac}
1+\frac{4\left[1-\cosh(k_x L_y))\right]}{k_x L_y\left[k_x L_y+\sinh(k_x L_y)\right]},
\end{equation}
the theoretical prediction is further improved at small $k_x$ as shown in Fig.~\ref{fig_dimer_giant_fluct}~(b). 
We note, however, that this factor is obtained for impermeable walls and the resulting correction is not exact for our reservoir boundaries.
For the reactive cases, the validity of \eqref{Sneqww} is questionable due to the nonlinear concentration gradients.
In fact, how to estimate the value of $h^2$ is not obvious.
Considering that $S^\mathrm{neq}_{w_1,w_1}$ is an averaged structure factor for different values of $y$, we estimate the value of $h^2$ from the profile of $\bar{w}_1(y)$ using a spatial average,
\begin{equation}
\label{hsq}
h^2= \frac{1}{L_y} \int_0^{L_y} \left(\frac{d\bar{w}_1}{dy}\right)^2 dy.
\end{equation}
Theoretical results obtained from \eqref{Sneqww}, \eqref{confinefac}, and \eqref{hsq} are shown in Fig.~\ref{fig_dimer_giant_fluct}~(b).
Remarkably, the $k^{-4}$ power law region is accurately predicted.
However, the theoretical prediction overestimates $S^\mathrm{neq}_{w_1,w_1}$ at small $k_x$ by several orders of magnitude for the reactive cases.
This is expected since the local linear gradient approximation eventually fails at large length scales.
The FHD equations linearized around a nonlinear stationary profile were studied in \cite{BedeauxZaratePagonabarragaSengersKjelstrup2011}; however, an explicit result for the static structure factor that we could compare with our numerical result was not obtained.

\subsection{\label{subsec_fingering}Fingering Instability with a Neutralization Reaction}

In this section we examine the development of asymmetric fingering patterns arising from a diffusion-driven gravitational instability in the presence of a neutralization reaction. 
We perform three-dimensional large-scale simulations of a double-diffusive instability occurring during the mixing of HCl and NaOH solution layers in a vertical Hele-Shaw cell (two parallel glass plates separated by a narrow gap).
This system has been studied experimentally and theoretically using a two-dimensional Darcy advection-diffusion-reaction model~\cite{LemaigreBudroniRiolfoGrosfilsWit2013, AlmarchaTrevelyanGrosfilsWit2013}.
Thermal fluctuations may play a key role in triggering the instability.
To the best of our knowledge, our simulations are the first ones to use a three-dimensional model and the first to include thermal fluctuations.
We investigate the effects of each stochastic component (mass flux, momentum flux, and chemistry) on the evolution of the system.
We initialize our simulations with natural mass and momentum fluctuations without any artificial perturbation, and therefore our simulation can be regarded as an \emph{ideal} experiment.

\subsubsection{Model Setup}

For the model setup and physical parameters, we follow the experiment of Lemaigre et al.~\cite{LemaigreBudroniRiolfoGrosfilsWit2013}.
We use cgs units unless otherwise specified and assume $T=293$ and atmospheric pressure.
The isothermal approximation has been justified by a linear stability analysis showing that the heat generated by the neutralization reaction
\begin{equation}
\label{neutralization_rxn}
\mathrm{HCl}+\mathrm{NaOH}\rightarrow\mathrm{NaCl}+\mathrm{H_2O}
\end{equation}
plays a negligible role in this problem~\cite{AlmarchaTrevelyanGrosfilsWit2013}.
We consider a Hele-Shaw cell with side lengths $L_x=L_y=1.6$ and $L_z=0.05$, with the $y$-axis pointing in the vertical direction, and the $z$-axis being perpendicular to the glass plates.
The domain is divided into grid cells with grid spacing $\D{x}=\D{y}=\D{z}=\num{6.25e-3}$ so there are $256\times256\times8$ cells.
We impose periodic boundary conditions in the $x$-direction and no-slip walls in the $z$-direction.
In the $y$-direction, we impose free-slip reservoir boundary conditions with concentrations that match the initial conditions of each layer.

We start with a gravitationally stable initial configuration, where an aqueous solution of NaOH with molarity 0.4~mol/L is placed on top of a denser aqueous solution of HCl with molarity 1~mol/L.
Each reactant and product is treated as a single charge-neutral species, giving the four species HCl ($s=1$), NaOH ($s=2$), NaCl ($s=3$), and water ($s=4$).
Under the approximation that the solution density $\rho$ is linearly dependent on the solute concentrations~\cite{LemaigreBudroniRiolfoGrosfilsWit2013}, the buoyancy force is expressed as
\begin{equation}
\V{f}(\V{w})= -\rho_0 \left(\sum_{s=1}^3 \frac{\alpha_s}{M_s} w_s\right) g \V{e}_y,
\end{equation}
where $\alpha_s$ is the solutal expansion coefficient, and $M_s$ is the molecular weight (in g/mol) of solute $s$.
We set $g=981$, $\rho_0=1$, and $\eta=0.01$.
The initial density difference between the two layers is approximately \num{4e-4}.
The Maxwell--Stefan binary diffusion coefficients are determined using \eqref{MSdiffcoeff} from the known trace diffusion coefficients of the solutes and the self-diffusion coefficient of water.
The values of $\alpha_s$ and the trace diffusion coefficients are obtained from Table~II in \cite{LemaigreBudroniRiolfoGrosfilsWit2013}.

Since the neutralization equilibrium lies far to the product side, we only consider the forward reaction.
We assume that the rate is given by the traditional LMA for a dilute solution, $a^+ = k n_1 n_2$.
However, we note that neutralization is a diffusion-limited reaction.
In other words, reaction occurs extremely fast (with rate $\lambda \sim 10^{11}\: \mathrm{s}^{-1}$), as soon as reactants encounter each other.
Because of this, the validity of the local LMA is questionable~\cite{DonevYangKim2018}.
The estimated value of $k \sim 10^{-11}\: \mathrm{cm^3 s^{-1}}$ is impractically large (converted using (31) in \cite{ErbanChapman2009}), and would require an unreasonably small $\D{t}$ for numerical stability.
For our simulations, we choose a smaller value $k=10^{-18}$ and $\D{t}=10^{-3}$ based on a deterministic numerical study presented in Appendix~\ref{appendix_fingering_kdt}.

The initial mass fractions in each cell are generated as the sum of mean values $\V{w}^0$ and natural fluctuations $\d{\V{w}}$. 
The mean mass fractions $\V{w}^0$ are set to $\V{w}^{0,\mathrm{upper}}=(0,0.0157,0,0.9843)$ in the upper half-domain and $\V{w}^{0,\mathrm{lower}}=(0.0358,0,0,0.9642)$ in the lower half-domain.
Assuming that natural fluctuations are Gaussian, we sample them using the known equilibrium structure factor at the mean state (Eq.~(D4) in \cite{DonevNonakaBhattacharjeeGarciaBell2015}),
\begin{equation}
\d{\V{w}}=\frac{1}{\sqrt{\rho_0\D{V}}}(\M{I}-\V{w}^0\V{1}^\mathrm{T})(\M{W}^0\M{M})^\frac12\V{z}^\text{mass},
\end{equation}
where $\M{W}^0=\mathrm{diag}(w_s^0)$, $\M{M}=\mathrm{diag}(m_s)$, and $\V{z}^\text{mass}$ is a vector with i.i.d.\ standard normal random variables.
The initial momentum fluctuations are generated in a similar manner,
\begin{equation}
\d{\V{v}}=\sqrt{\frac{k_\mathrm{B}T}{\rho_0\D{V}}}\V{z}^\text{mom},
\end{equation}
where $\V{z}^\text{mom}$ is a vector with i.i.d.\ standard normal random variables, followed by an $L^2$ projection onto the space of divergence-free vector fields.

We use the Langevin chemistry description given in \eqref{OmegaLangevin} and the BDS advection scheme.
We can justify the use of the CLE by noting that the system is near the macroscopic limit because fluctuations are weak.
For centered advection, we observe oscillations around the interface of fingers (not shown) for the chosen grid spacing as expected due to the high cell P\'eclet number.
When the grid spacing is reduced to half, oscillations become less pronounced without changing the results significantly (not shown).

\subsubsection{Effects of Thermal Fluctuations}

\begin{table}
\begin{center}
\small
\caption{\label{table_sim}
Four simulations performed for the buoyancy-driven instability with an acid-base neutralization reaction.}
\begin{tabular}{|c|c|c|c|c|c|}
\hline
\multirow{2}{*}{}   & \multirow{2}{*}{chemistry} & \multicolumn{2}{c|}{initial fluctuations} & \multicolumn{2}{c|}{stochastic fluxes} \\
\cline{3-6}
                    &               & mass  & momentum  & mass  & momentum  \\
\hline
simulation A        & stochastic    & on    & on        & on    & on        \\
\hline
simulation B        & no reaction   & on    & on        & on    & on        \\
\hline
simulation C        & deterministic & off   & on        & off   & on        \\
\hline
simulation D        & deterministic & off   & off       & off   & on        \\
\hline
\end{tabular}
\end{center}
\end{table} 

We perform four FHD simulations changing how chemistry is treated, whether natural mass/momentum fluctuations are initially imposed, and whether subsequently stochastic mass/momentum fluxes are included, as summarized in Table~\ref{table_sim}.

By comparing the results of simulations~A, B, C, and D, we can assess the effects of chemo-hydrodynamic coupling and thermal fluctuations on the fingering pattern formation.
For a perfectly flat initial interface, thermal fluctuations play an essential role in perturbing the interface at early times, but once an uneven interface appears, the dynamic instability dominates and thermal fluctuations are expected to play a secondary role in subsequent pattern formation, as we previously confirmed in the absence of reactions~\cite{DonevNonakaBhattacharjeeGarciaBell2015}.

\begin{figure}
\includegraphics[width=\textwidth]{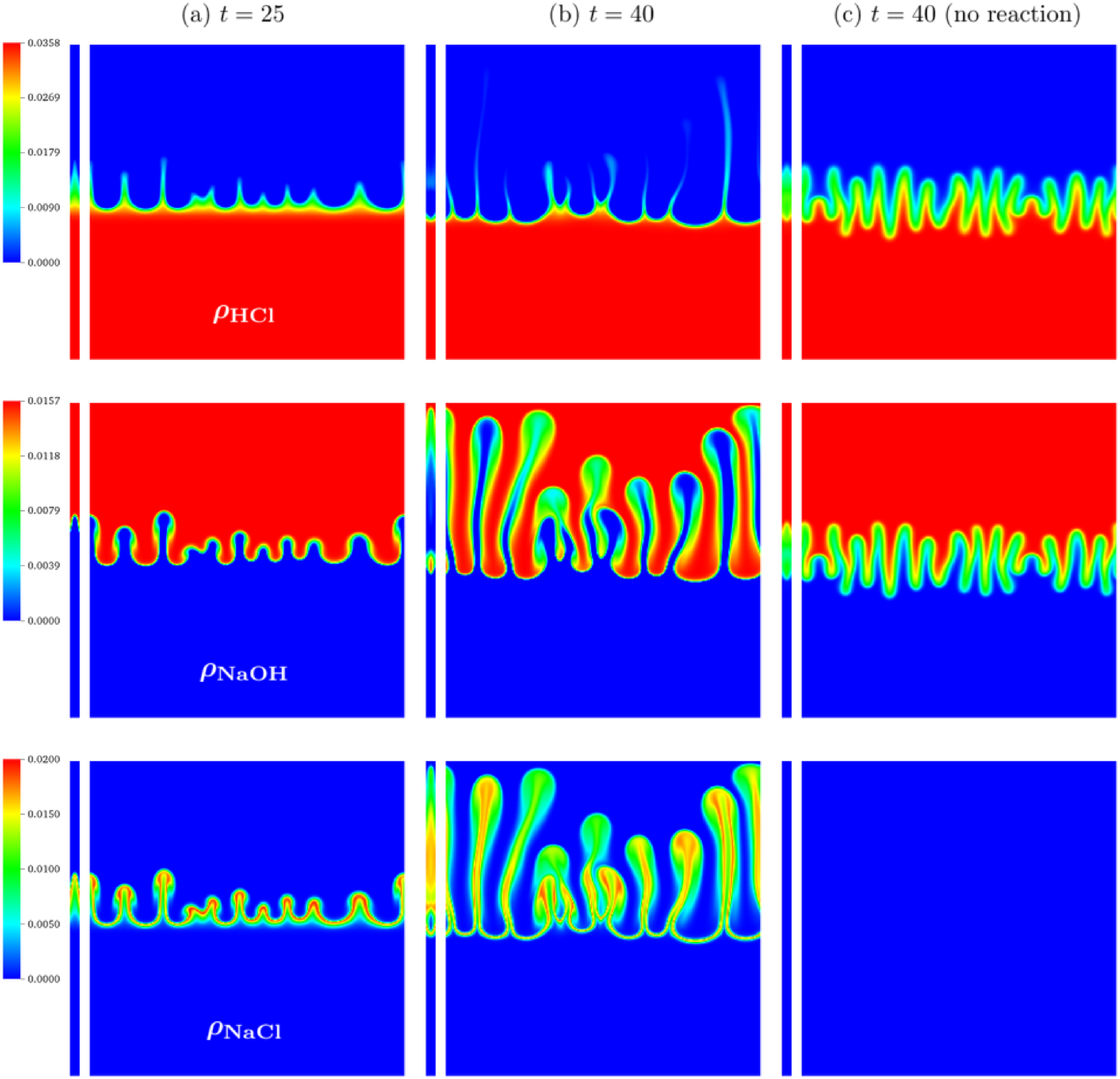}
\caption{\label{fig_ABF_FHD}
Asymmetric growth of convective chemo-hydrodynamic fingering patterns in a Hele-Shaw cell, induced by a double-diffusive instability in the presence of a neutralization reaction.
The left and middle columns, (a) and (b), depict the mass density profiles of chemical species at $t=25$ and $t=40$ (simulation~A), whereas the right column~(c) displays the non-reactive case at $t=40$ (simulation B).
The upper, mid, and bottom rows show $\rho_\mathrm{HCl}$, $\rho_\mathrm{NaOH}$, and $\rho_\mathrm{NaCl}$, respectively.
For each species, two-dimensional slices of the three-dimensional field $\rho_s(x,y,z)$ are shown.
The square images show $\rho_s(x,y,z=L_z/2)$ (halfway between the glass plates) and the thin vertical side bars show the slice $\rho_s(x=0,y,z)$ corresponding to the left edge of the square images.
Both simulations were initiated with natural mass and momentum fluctuations without any artificial perturbation.}
\end{figure}

We compare the reactive case (simulation~A) with the non-reactive case (simulation~B) in Figure~\ref{fig_ABF_FHD}.
As also seen in the experiment~\cite{LemaigreBudroniRiolfoGrosfilsWit2013}, an asymmetric growth of fingers is observed in the reactive case.
In addition, the growth of fingers is much faster when the reaction is present.
This is due to the coupling of the fast neutralization reaction and the fast diffusion of the acid species.
Disparities between the acid and base species can be also seen in the concentration profiles around the fingers; such disparities are not observed in the non-reactive case.
We point out that the concentration develops three-dimensional profiles that are not constant across the thickness of the Hele-Shaw cell, as can be seen from the side bars ($y-z$ cross-sections) in the figure. 
Such structure would not be captured by the two-dimensional Darcy approximation used in prior computational studies~\cite{LemaigreBudroniRiolfoGrosfilsWit2013, AlmarchaTrevelyanGrosfilsWit2013}. 

\begin{figure}
\includegraphics[width=\textwidth]{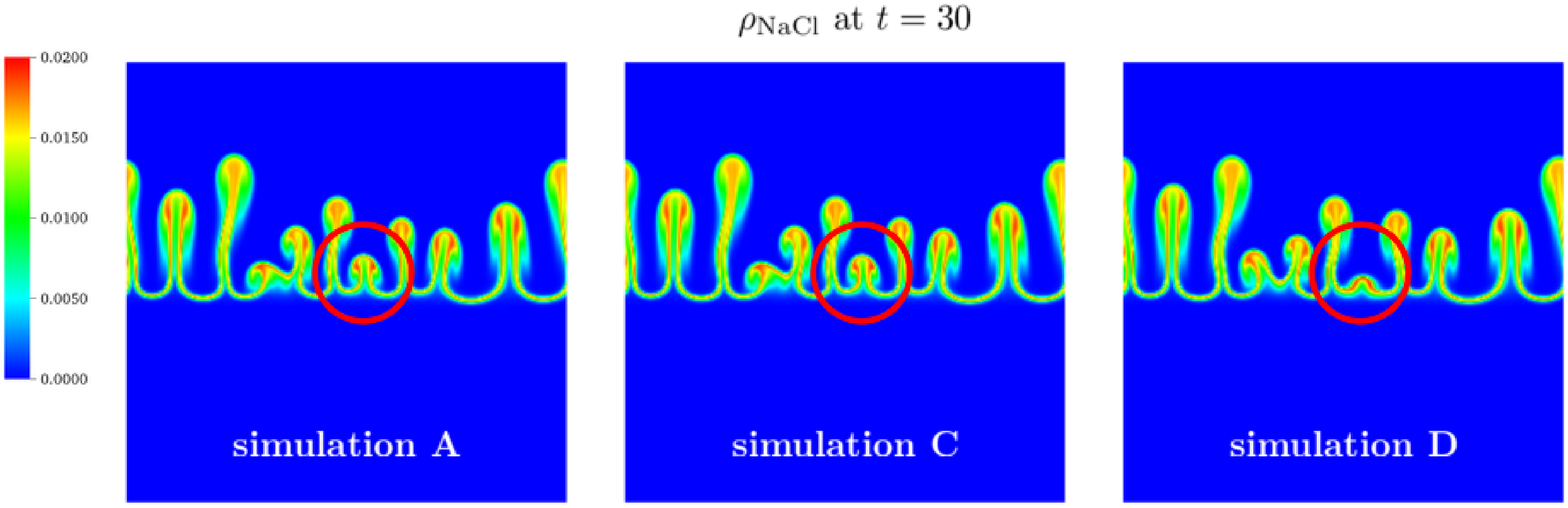}
\caption{\label{fig_ABF_FHD_comp}
Influence of different types of thermal fluctuations on the formation of fingering patterns.
We compare the mass density profiles  $\rho_s(x,y,z=L_z/2)$ of NaCl at $t=30$ halfway between the glass plates for three simulations.
Simulation~A (left) corresponds to the full fluctuating hydrodynamics equations.
Compared with simulation~A, all stochastic mass components (stochastic mass flux and stochastic chemistry) are omitted in simulation~C (middle).
Simulation~D (right) is similar to simulation~C but with initial velocity fluctuations also removed. 
Red circles indicate areas with the biggest differences.}
\end{figure}

In Fig.~\ref{fig_ABF_FHD_comp}, we compare simulations~C and D with simulation~A to investigate the contribution of each stochastic component.
Compared with the full fluctuating hydrodynamics (simulation~A), all stochastic \emph{mass} components are omitted in simulation~C.
However, the resulting fingering patterns are essentially the same as in simulation~A.
This indicates that contributions of stochastic mass fluxes and stochastic chemistry are negligible in this example.
Instead, concentration fluctuations driven by the stochastic momentum flux dominate the formation of an uneven interface. 
This can be confirmed by the comparison of simulation~A with simulation~D, where initial velocity fluctuations are turned off compared with simulation~C, and only stochastic momentum fluxes are included.
The resulting fingering patterns are virtually the same with slight differences caused by differences in the initial velocity conditions.
This is consistent with the fact that any initial momentum conditions are quickly damped out in a liquid with a high Schmidt number.
In fact, in a simulation similar to simulation~C but \emph{without} subsequent stochastic momentum fluxes, it takes more time ($\sim$ 5~s longer) for fingering patterns to start to grow.
Hence, we confirm that velocity fluctuations driving giant composition fluctuations dominate the triggering of the instability starting from a perfectly flat interface.

It is important to note that our simulation results show that the thermal fluctuations are sufficiently large to kick off the instability on a time scale comparable to that when a macroscopic initial perturbation is imposed.
The fingering patterns observed in simulation~A at $t=40\;\mathrm{s}$ are quite comparable to the experimental result shown in Figure~1~(e) in \cite{LemaigreBudroniRiolfoGrosfilsWit2013} at $t=30\;\mathrm{s}$. 
Of course, in the actual experiments the initial interface is not perfectly flat due to imperfections in the preparation of the initial configuration.

\section{\label{sec_concl}Summary and Discussion}

We have developed a fluctuating hydrodynamics (FHD) formulation and numerical methodology for stochastic simulation of reactive liquid mixtures.
Our approach robustly models a wide range of microliquids, including dilute solutions as well as mixtures with no single dominant solvent.
Our multispecies transport model is based on Maxwell--Stefan cross-diffusion, incorporates a stochastic chemistry description based on the chemical master equation (CME), and couples reaction-diffusion with a stochastic Navier--Stokes equation for the fluid velocity.
Our numerical method is based on several techniques that helped us gain computational efficiency without compromising accuracy.
Specifically, the implicit treatment of momentum dissipation allowed us to avoid the severe restriction on time step size imposed by the small Reynolds number and large Schmidt number.
The use of the tau leaping method enabled us to sample CME-based chemistry efficiently while correctly sampling large deviations from chemical equilibrium~\cite{KellyVandenEijnden2017}.
For a binary liquid mixture undergoing a dimerization reaction, we demonstrated the thermodynamic consistency of our overall formulation beyond the Gaussian approximation, and accurately reproduced the equilibrium Einstein distribution for both dilute solutions and liquid mixtures. 
Owing to a careful treatment of strong concentration fluctuations and vanishing species, our numerical method remained robust even for cells with as few as ten molecules; coarse-graining at such small scales is at the limits of fluctuating hydrodynamics.

Our numerical results for the spectrum of giant nonequilibrium fluctuations in a binary mixture undergoing a dimerization reaction were not in good agreement with theoretical predictions for smaller wavenumbers. 
We believe that this mismatch is due to the fact that we used a very simple theory that assumes the gradient is constant and weak. 
A more accurate theory requires linearizing the FHD equations around the nonlinear steady-state solution of the macroscopic equations, and proper treatment of the boundary conditions. 
Such a linearization was carried out in~\cite{BedeauxZaratePagonabarragaSengersKjelstrup2011} without accounting for the boundary conditions (see in particular Eq.~(15) in~\cite{BedeauxZaratePagonabarragaSengersKjelstrup2011}).
Nevertheless, analytical computation of the structure factor proved too difficult and the authors used a perturbative analysis for which the zeroth-order approximation is the simple approximation that the applied gradient is constant and weak and the system is everywhere near chemical equilibrium.
An explicit formula for the next-order correction was not obtained for the static structure factor.
Our computations showed that the simple zeroth-order theory, while giving a qualitatively correct picture of how reactions affect giant fluctuations, overestimates the fluctuations by orders of magnitude at small wavenumbers.

We performed the first three-dimensional simulations of a buoyancy-driven instability in the presence of an acid-base neutralization reaction. 
Our results demonstrate that velocity fluctuations generate giant concentration fluctuations that are sufficiently large to drive the initial growth of the instability, even when the initial interface is perfectly flat. 
In particular, we found that thermal fluctuations can trigger the instability on a time scale comparable to that observed in recent experiments, although a direct comparison is not possible because the exact initial condition in experiments is hard to control or measure.

In our prior work on reaction-diffusion systems~\cite{KimNonakaBellGarciaDonev2017}, we treated diffusion implicitly.
This allowed us to use time step sizes an order of magnitude larger than the stability limit imposed by species diffusion.
In this work we treated diffusion explicitly because momentum diffusion is much faster than mass diffusion in liquids, and thus the time step size was primarily limited by the requirement to accurately resolve momentum dynamics at small scales.
Nevertheless, in a number of problems, such as, for example, catalytic micropumps~\cite{EsplandiuFarniyaReguera2016} or electroconvective instabilities at large applied voltages~\cite{AndersenWangSchiffbauerMani2017}, the time scales of interest are those at which diffusion reaches a quasi-steady state in at least one direction.
In this case one must treat diffusion implicitly. 
This is straightforward in principle but requires the development of several nontrivial components. 
First, because the diffusion of all species is coupled in generic mixtures, one must develop either temporal integrators that treat only the diagonal part of the diffusion matrix implicity, or develop a multispecies multigrid solver for coupled implicit discretizations. 
Second, the semi-implicit temporal integrators developed in \cite{KimNonakaBellGarciaDonev2017} must be modified to integrate the momentum equation in a way that is robust for large Schmidt numbers. 
Third, boundary conditions need to be handled, both in the diffusion solver, and in the coupling between diffusion and advection for reservoir boundaries.

In this work we assumed the validity of a Boussinesq approximation, neglecting the change in density with composition at a given pressure and temperature, as dictated by the equation of state (EOS) of the mixture. 
This is a limiting approximation in practice, especially for reactive mixtures in which reactions can rapidly change density locally. 
In prior work~\cite{DonevNonakaBhattacharjeeGarciaBell2015}, we accounted for the density dependence on composition using low Mach asymptotics. 
It is important to observe that the multispecies low Mach model proposed in \cite{DonevNonakaBhattacharjeeGarciaBell2015} applies even to ideal gas mixtures, not just liquid mixtures. 
There are several difficulties with extending the formulation and algorithms we developed in prior work to reactive low Mach number models. 
First, reactions can lead to local changes in pressure which, in the low Mach limit, must get instantaneously distributed throughout the system as a global adjustment of the background thermodynamic pressure. 
It is anticipated that barodiffusion will have to be accounted for to achieve thermodynamic consistency when the chemical potentials depend nontrivially on pressures.
Second, enforcing the EOS will require a nonlinear iteration of a coupled mass-momentum diffusion system, unlike the simpler case considered in \cite{DonevNonakaBhattacharjeeGarciaBell2015} where we could enforce a linear EOS with only a decoupled linear Stokes solve. Both of these difficulties are compounded if one wishes to treat diffusion implicitly or to account for energy transport in a non-isothermal generalization.

In future work, we will account for electrochemistry by incorporating charged species into our model, similar to the developments in \cite{PeraudNonakaChaudhriBellDonevGarcia2016} for the non-reactive case.
By using electroneutral asymptotics~\cite{GriffithPeskin2013}, we will be able to model the species (HCl, NaOH, and NaCl) in the acid-base fingering instability as separate ions ($\mathrm{H}^+$, $\mathrm{OH}^-$, $\mathrm{Na}^+$, and $\mathrm{Cl}^-$), which is physically correct given that HCl and NaOH are both strong electrolytes. 
Resolving the diffusion of each ion individually is required to correctly model electrodiffusion in mixtures with more than two ions.
Incorporating charged species will also allow us to simulate weak electrolyte solutions (in which molecules do not fully disassociate into ions), catalytic motors~\cite{EsplandiuFarniyaReguera2016}, and electrokinetic locomotion~\cite{Moran2011, MoranPosner2017}.

\begin{acknowledgments}
We would like to thank Anne De Wit for helpful discussions regarding gravitational instabilities in the presence of neutralization reactions, and thank Eric Vanden-Eijnden for discussions regarding tau leaping and large deviation theory.
This work was supported by the U.S.~Department of Energy, Office of Science, Office of Advanced Scientific Computing Research, Applied Mathematics Program under contract DE-AC02-05CH11231 and Award Number DE-SC0008271.
This research used resources of the National Energy Research Scientific Computing Center, a DOE Office of Science User Facility supported by the Office of Science of the U.S.~Department of Energy under Contract No.~DE-AC02-05CH11231.
\end{acknowledgments}

\appendix

\section{\label{appendix_Wchi}Diffusion Matrix with Vanishing Species}

In this appendix we derive the analytic expressions~\eqref{eq:Wchi} for the matrix $\M{W}\M{\chi}$ in the limit of vanishing species.
For simplicity, we consider a case where the last species among $N$ species vanishes:
\begin{subequations}
\begin{align}
&w_N \rightarrow 0^+,\\
&w_i \rightarrow w_i^0>0 \quad (i=1,\dots,N-1) \mbox{ with }\sum_{i=1}^{N-1} w_i^0 = 1.
\end{align}
\end{subequations}
We show next that each component of $\M{W}\M{\chi}$ converges to
\begin{subequations}
\label{Wchi_appendix}
\begin{align}
\label{Wchi_ij}
&(\M{W}\M{\chi})_{ij} \rightarrow w_i^0 \chi^\mathrm{sub}_{ij} & (i,j&=1,\dots,N-1)\\
\label{Wchi_NN}
&(\M{W}\M{\chi})_{NN} \rightarrow \frac{m_N D_N}{\bar{m}^0} & & \\
\label{Wchi_Ni}
&(\M{W}\M{\chi})_{Ni} \rightarrow 0 & (i&=1,\dots,N-1) \\
\label{Wchi_iN}
&(\M{W}\M{\chi})_{iN} \rightarrow w_i^0 D_N \left[\sum_{k=1}^{N-1}\frac{\chi_{ik}^\mathrm{sub} x_k^0}{\textsc{\DJ}_{kN}}-\frac{m_N}{\bar{m}^0} \right] & (i&=1,\dots,N-1)
\end{align}
\end{subequations}
where $\M{\chi}^\mathrm{sub}$ is the diffusion matrix of the subsystem consisting of non-vanishing species with $\V{w}^0=(w_1^0,\dots,w_{N-1}^0)$, and $\V{x}^0$ and $\bar{m}^0$ are computed from \eqref{w2x} and \eqref{barm} using $\V{w}^0$.
We also show that the trace diffusion coefficient $D_N$ of species $N$ in the fluid mixture with composition $\V{w}^0$ is expressed as
\begin{equation}
\label{D_N}
D_N = \left[ \sum_{k=1}^{N-1}\frac{x_k^0}{\textsc{\DJ}_{kN}}\right]^{-1}.
\end{equation}

By rearranging the definition of $\M{\chi}$~\cite{DonevNonakaBhattacharjeeGarciaBell2015},
\begin{equation}
\M{\chi} = (\M{\Lambda}+\alpha\V{w}\V{w}^\mathrm{T})^{-1}-\frac{1}{\alpha}\V{1}\V{1}^\mathrm{T},
\end{equation}
where $\alpha\ne 0$, and using $\V{1}^\mathrm{T}\M{\Lambda}=\V{0}$, $\M{\chi}\V{w}=\V{0}$, and $\V{1}^\mathrm{T}\V{w}=1$, we obtain
\begin{equation}
\label{chiLambda}
\M{\chi}\M{\Lambda}+\V{1}\V{w}^\mathrm{T} = \M{I}.
\end{equation}
Looking at the $(N,N)$-component of \eqref{chiLambda} and using \eqref{Lambda}, we have
\begin{equation}
\label{NNcomp}
x_N\: \chi_{NN} \sum_{i=1}^{N-1} \frac{x_i}{\textsc{\DJ}_{iN}}\left(1-\frac{\chi_{Ni}}{\chi_{NN}}\right)+w_N = 1.
\end{equation}
Noting that $\chi_{NN}=O(w_N^{-1})$ and
\begin{equation}
\label{chiNichiNN}
\frac{\chi_{Ni}}{\chi_{NN}} \rightarrow 0,
\end{equation}
we obtain \eqref{Wchi_NN} and \eqref{D_N} by taking the limit of \eqref{NNcomp} using \eqref{w2x}. 
Then \eqref{Wchi_NN} and \eqref{chiNichiNN} imply \eqref{Wchi_Ni}.
Applying the same technique to the $(i,N)$-component of \eqref{chiLambda} for $i=1,\dots,N-1$, we obtain \eqref{Wchi_iN}.

We also observe that \eqref{Wchi_appendix} gives
\begin{equation}
\label{barF_N}
\overline{\V{F}}_N = -\rho_0 \frac{m_N D_N}{\bar{m}^0}\grad x_N.
\end{equation}
Hence, the diffusion of the dilute species becomes decoupled from those of other species and its trace diffusion coefficient is given by \eqref{D_N}.

\section{\label{appendix_fingering_kdt}Rate Constant of Neutralization Reaction}

In this appendix we determine an appropriate value for the rate constant $k$ of the neutralization reaction~\eqref{neutralization_rxn} for the simulations of the fingering example reported in Section~\ref{subsec_fingering}.
As mentioned in the main text, the estimated value of $k\sim 10^{-11}$ is too large, as it requires impractically small time step sizes.
By performing deterministic simulations, we investigate a range of values for $k$ to determine at what point increasing $k$ stops changing the results.
We also examine the convergence of the results using different time step sizes.

For these deterministic simulations, we use a smaller domain (half the length in the $x$- and $y$-directions) with the same grid spacing.
To generate an initial configuration with an uneven interface, we introduce random perturbations of composition in each cell immediately above the interface and set
\begin{equation}
w_s^0 = a U w_s^{0,\textrm{lower}} + (1-a U) w_s^{0,\textrm{upper}},
\end{equation}
where $a = 0.1$ and $U$ is a standard normal random number generated independently in each cell.
We compute fingering patterns for several values of $k$ from $10^{-23}$ to $10^{-15}$, with several values of $\D{t}$ ranging from $10^{-3}$ to $10^{-2}$, using the \emph{same} random initial configuration.
To assess the similarity of two simulation results, we compute the gross NaCl production $\rho_0 \int w_3(\V{r},t) d\V{r}$, as well as the $L^1$-norm of the $v_y$ field $\lVert v_y \rVert = \int \lvert v_y(\V{r},t)\rvert d\V{r}$.

\begin{figure}
\centerfloat
\includegraphics[width=1.2\textwidth]{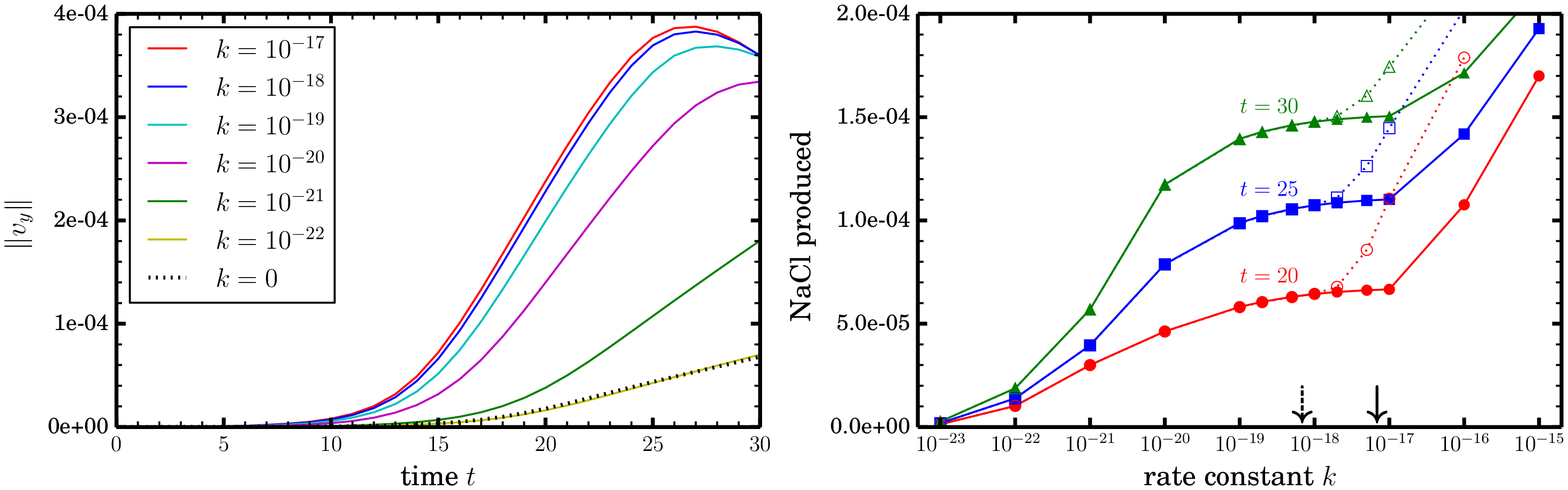}
\caption{\label{fig_ABF_det}
Effects of the reaction rate constant $k$ on the fingering instability observed when a layer of NaOH is placed on top of HCl solution, for deterministic simulations with a randomly perturbed initial interface.
Panel~(a) shows the time profiles of the norm of the $v_y$ field for various values of $k$.
Panel~(b) shows gross NaCl production up to time $t$ as a function of $k$.
Solid lines denote the results for $\D{t}=10^{-3}$, whereas dotted lines in the same colors depict the results for $\D{t}=10^{-2}$.
Arrows indicate $k=4/(n^0_\mathrm{HCl}\D{t})$ for $\D{t}=10^{-2}$ (dotted line) and $\D{t}=10^{-3}$ (solid line), where $n^0_\mathrm{HCl}$ is the initial number density of HCl in the lower layer.}
\end{figure}

Figure~\ref{fig_ABF_det}~(a) shows the time evolution of $\lVert v_y \rVert$ for various values of $k$ for $\D{t}=10^{-3}$.
As $k$ increases, $\lVert v_y \rVert$ grows faster, indicating that fingers grow faster.
For $10^{-22} \lesssim k \lesssim 10^{-19}$, time profiles change significantly depending on the value of $k$.
On the other hand, for $k \gtrsim 10^{-19}$, the change becomes less significant.
Also, time profiles for $k\lesssim 10^{-22}$ coincide with that of the non-reactive case.
This suggests that there are three different regimes for $k$: slow, intermediate, and fast reaction regimes.
The gross NaCl production shown in Fig.~\ref{fig_ABF_det}~(b) exhibits similar behavior.
While more NaCl is produced as $k$ increases, the growth slows down around $k\approx 10^{-19}$ and a plateau is observed beyond this value.
Hence, from a modeling point of view, one can simulate the neutralization reaction using a value of $k$ from the plateau region.
It is important to note, however, that one cannot choose an arbitrarily large value of $k$ due to the stability limit imposed by our explicit tau-leaping treatment of reactions.
In fact, fingering patterns obtained using $\D{t}=10^{-2}$ and $10^{-3}$ (not shown) are essentially the same for $k\lesssim 10^{-18}$.
However, both results start to show unphysical behaviors for $k\D{t} > 4/n^0_\mathrm{HCl}$, where $n^0_\mathrm{HCl}$ is the initial number density of HCl species in the lower layer, as can be seen from the abrupt increase of the gross NaCl production in Fig.~\ref{fig_ABF_det}~(b).

Based on these observations, we choose $k=10^{-18}$ and $\D{t}=10^{-3}$.
The value of $\D{t}$ is much smaller than the mass diffusion stability limit.
As shown in Fig.~\ref{fig_ABF_det}~(b), $\D{t}\lesssim 10^{-2}$ is required to guarantee stability when the reaction is stiff and $k \approx 10^{-18}$.
It is noted, however, that $\D{t}\lesssim 10^{-3}$ is required to give a reasonable CFL number for momentum diffusion $\nu\D{t}/\D{x}^2 = 0.256$.
This is because small time-integration errors in the velocity field at early times can cause significant perturbations at later times because of the growing instability.
If the exact time evolution at early times is not important, one can safely use $\D{t}=10^{-2}$ without sacrificing physical fidelity.

\bibliography{BousqReactFHD.bbl}

\end{document}